\documentclass[twoside]{article}
\usepackage{qic,epsfig}

\usepackage{amsmath}
\usepackage{latexsym}
\usepackage{bm,bbm}
\usepackage{graphicx}

\usepackage[matrix,frame,arrow]{xy}
\usepackage{amsmath}

\newcommand{\ket}[1]{\left\vert{#1}\right\rangle}
\newcommand{\qw}[1][-1]{\ar @{-} [0,#1]}
\newcommand{\qwx}[1][-1]{\ar @{-} [#1,0]}

\newcommand{\gate}[1]{*{\xy *+<.6em>{#1};p\save+LU;+RU **\dir{-}\restore\save+RU;+RD **\dir{-}\restore\save+RD;+LD **\dir{-}\restore\POS+LD;+LU **\dir{-}\endxy} \qw}

\newcommand{\measureD}[1]{*{\xy*+=+<.5em>{\vphantom{\rule{0em}{.1em}#1}}*\cir{r_l};p\save*!R{#1} \restore\save+UC;+UC-<.5em,0em>*!R{\hphantom{#1}}+L **\dir{-} \restore\save+DC;+DC-<.5em,0em>*!R{\hphantom{#1}}+L **\dir{-} \restore\POS+UC-<.5em,0em>*!R{\hphantom{#1}}+L;+DC-<.5em,0em>*!R{\hphantom{#1}}+L **\dir{-} \endxy} \qw}

\newcommand{\control}{*!<0em,.025em>-=-{\bullet}}

\newcommand{\ctrl}[1]{\control \qwx[#1] \qw}

\newcommand{\targ}{*!<0em,.019em>=<.79em,.68em>{\xy {<0em,0em>*{} \ar @{ - } +<.4em,0em> \ar @{ - } -<.4em,0em> \ar @{ - } +<0em,.36em> \ar @{ - } -<0em,.36em>},<0em,-.019em>*+<.8em>\frm{o}\endxy} \qw}
\newcommand{\qswap}{*=<0em>{\times} \qw}
\newcommand{\dstick}[1]{*!U!<0em,.5em>=<0em>{#1}}
\newcommand{\Qcircuit}[1][0em]{\xymatrix @*[o] @*=<#1>}

\newcommand{\tr}{\mathrm{tr}}
\renewcommand{\1}{\ensuremath{\mathbbm I}}
\newcommand{\Cplx}{\ensuremath{\mathbbm C}}
\newcommand{\Real}{\ensuremath{\mathbbm R}}
\newcommand{\diag}{\mathrm{diag}}
\newtheorem{fact}{Fact}
\newtheorem{prop}{Proposition}

\newcommand{\halmos}{}
\newcommand{\proofof}[2]{\noindent {\bf Proof of #1.} #2 $\Box$.}
\newcommand{\scalar}[2]{\langle #1 | #2 \rangle}

\newcommand{\states}{\Omega}

\begin{document}
\setlength{\textheight}{8.0truein}    

\runninghead{Sub-- and super--fidelity as bounds for quantum fidelity}
            {J. A. Miszczak, Z. Pucha{\l}a, P. Horodecki, A. Uhlmann, and K. Zyczkowski}

\normalsize\textlineskip
\thispagestyle{empty}
\setcounter{page}{103}

\copyrightheading{9}{1\&2}{2009}{0103--0130}

\vspace*{0.68truein}

\alphfootnote

\fpage{103}

\centerline{\bf
SUB-- AND SUPER--FIDELITY AS BOUNDS FOR QUANTUM FIDELITY}
\vspace*{0.37truein}
\centerline{\footnotesize
JAROS{\L}AW ADAM MISZCZAK~~~~ZBIGNIEW PUCHA{\L}A}
\vspace*{0.015truein}
\centerline{\footnotesize\it Institute of Theoretical and Applied Informatics, Polish Academy
of Sciences,}
\baselineskip=10pt
\centerline{\footnotesize\it Ba{\l}tycka 5, 44-100 Gliwice, Poland}
\vspace*{10pt}
%

\centerline{\footnotesize
PAWE{\L} HORODECKI}
\vspace*{0.015truein}
\centerline{\footnotesize\it
Faculty of Applied Physics and Mathematics,
Gda{\'n}sk University of Technology,
}
\baselineskip=10pt
\centerline{\footnotesize\it Narutowicza 11/12, 80-952 Gda{\'n}sk, Poland}
\baselineskip=10pt
\centerline{\footnotesize\it and}
\baselineskip=10pt
\centerline{\footnotesize\it
National Quantum Information Centre of Gda{\'n}sk, 
}
\baselineskip=10pt
\centerline{\footnotesize\it Andersa 27, 81-824 Sopot, Poland}
\vspace*{10pt}

\centerline{\footnotesize
ARMIN UHLMANN}
\vspace*{0.015truein}
\centerline{\footnotesize\it Institute of  Theoretical Physics, University of Leipzig,}
\baselineskip=10pt
\centerline{\footnotesize\it Vor dem Hospitaltore 1, D-04103 Leipzig, Germany}
\vspace*{10pt}

\centerline{\footnotesize
KAROL \.ZYCZKOWSKI}
\vspace*{0.015truein}
\centerline{\footnotesize\it
Instytut Fizyki im. Smoluchowskiego, Uniwersytet
Jagiello{\'n}ski,
}
\baselineskip=10pt
\centerline{\footnotesize\it Reymonta 4, 30-059 Krak{\'o}w, Poland}
\baselineskip=10pt
\centerline{\footnotesize\it and}
\baselineskip=10pt
\centerline{\footnotesize\it
Centrum Fizyki Teoretycznej, Polska Akademia Nauk, 
}
\baselineskip=10pt
\centerline{\footnotesize\it Aleja Lotnik{\'o}w 32/44, 02-668 Warszawa, Poland}

\vspace*{0.225truein}

\publisher{May 19, 2008}{September 30, 2008}

\vspace*{0.21truein}

\abstracts{
  We derive several bounds on fidelity between quantum states.
  In particular we show that fidelity is bounded from above by
 a simple to compute quantity we call super--fidelity.
It is analogous to another quantity called sub--fidelity.  For any two states
 of a two--dimensional quantum system ($N=2$) all three quantities coincide.
We demonstrate that sub-- and super--fidelity are concave functions.
We also show that super--fidelity is super--multiplicative
 while sub--fidelity is sub--multiplicative and
design feasible schemes to measure these quantities in an experiment.
 Super--fidelity can be used to define a distance between
  quantum states. With respect to this metric the set
  of quantum states forms a part of a $N^2-1$ dimensional hypersphere.
}{}{}

\vspace*{10pt}

\vspace*{3pt}
\communicate{R Jozsa~\&~M Mosca }

\vspace*{1pt}\textlineskip    
\section{Introduction}

By processing quantum information we wish to transform a quantum state in a
controlled way.  Taking into account inevitable interaction with an environment
and possible imperfection of real dynamics it is then crucial to characterize
quantitatively, to what extend a given quantum state gets close to its target.
For this purpose one often uses {\sl fidelity}  \cite{Jo94}, here denoted by
$F$. That quantity has also been called {\sl transition probability} 
\cite{Uh76}: Operationally it is the maximal success probability of changing a
state to another one by a measurement in a larger quantum system. If both
quantum states are pure, fidelity is the squared
overlap between them.

In the general case fidelity between
any two mixed states is the function of
the trace norm of the product of their square roots.
Thus analytical evaluation of fidelity, or its direct
experimental measurement  becomes a cumbersome task.
Hence there is a need for other quantities, which
bound fidelity and are easier to compute and measure.

The aim of this work is to present some bounds for fidelity
and to develop experimental schemes to estimate  it 
for an arbitrary pair of mixed quantum states.
In particular we find an upper bound for fidelity by a simple
quantity which is the function of purity of both states and the
trace of their product.  Since it possesses some nice algebraic
properties we believe it may become useful in future research
and propose to call it {\sl super--fidelity}.
In a sense it is a quantity complementary
to the one forming the lower bound  proved in \cite{Uhlm00},
and we tend to call {\sl sub--fidelity}.
For any  two  one--qubit states all three quantities coincide.
Fidelity is well known to be multiplicative with respect to the tensor product.
In this work we prove that
super--fidelity is concave and super--multiplicative,
while  sub--fidelity  is concave and sub--multiplicative.

Fidelity can be used to define the Bures distance  between quantum states
and the Bures angle.
As shown by Uhlmann  in \cite{Uh92}
the Bures geometry of the set of one--qubit  states ($N=2$),
is equivalent to a three-dimensional  hemisphere $\frac{1}{2} S^3$.
The set of density operators, ${\states}_N$, becomes the space
of non-constant curvature by the Bures metric for $N\ge 3$,
\cite{Di99b}.

We construct distance and angle analogous to the Bures distance
out of super--fidelity in a similar way.
With respect to this metric the set  ${\states}_N$
forms a fragment of a $N^2-1$ dimensional hypersphere
with the maximally mixed state $\rho_*:={\1}/N$
at the pole.
A linear function of super--fidelity
was earlier used by Chen {\it et al.} \cite{CFUZ02}
to analyze the set of mixed quantum states
and demonstrate its hyperbolic interpretation.

This paper is organized as follows. In Section II the definition and basic
properties of fidelity are reviewed. Sections III and IV are devoted 
to bounds on fidelity.  In Section V we define sup-- and super--fidelity 
and investigate their properties. Experimental schemes designed to measure
these quantities are presented in section VI. In Section VII we
analyze the geometry of the set of quantum states
induced by the distance derived by super--fidelity.
Concluding remarks are followed by
appendices, in which we prove necessary
lemmas and present the collection of useful algebraic facts.

\section{Fidelity between quantum states}\label{sec:fid1}

Consider an $N$-- dimensional Hilbert space ${\cal H}_N$.
A linear  operator $\rho: {\cal H}_N \to {\cal H}_N$
represents quantum state if it is Hermitian,
semipositive,  $\rho=\rho^{\dagger}\ge 0$,
 and normalized, tr$\rho=1$.
Let ${\states}_N$ denote the set of all mixed quantum states
of size $N$.

Fidelity between quantum states $\rho_1$ and $\rho_2$ is defined as
\cite{Uh76,Jo94},
\begin{equation}
\label{fidel}
F(\rho_1, \rho_2) =  \left(\tr|\sqrt{\rho_1}\sqrt{\rho_2}| \right)^2
=  || \rho_1^{1/2} \rho_2^{1/2}||_1^2,
\end{equation}
where $||\cdot||_1$ is Schatten 1-norm (trace norm),
\begin{equation}
||A||_1 = \tr|A| := \tr\sqrt{AA^{\dagger}} .
\end{equation}
Alternatively, the  trace norm of an operator
 can be expressed as the sum of  its singular values,
$||A||_1 =\sum_{i=1}^n \sigma_i(A)$.
Here $\sigma_i(A)$ is equal to the
square root of the corresponding eigenvalue of the positive matrix
$AA^{\dagger}$  -- see e.g. \cite{bhatia}.

There are different uses of the name {\it fidelity}.
In \cite{Jo94} the older notion {\it transition probability}
has been renamed {\it fidelity} by Jozsa.
In \cite{NC00} $\sqrt{F}$ has been called {\it fidelity},
while \cite{BZ06} uses Jozsa's notion, and to the
latter convention we shall stick in calling fidelity the
expression in Eq. (\ref{fidel}).

For pure states the definition (\ref{fidel}) is reduced to the
transition probability.
If one state is pure, $\rho_{1}=|\psi\rangle \langle \psi|$,
then $F(\rho_1, \rho_2) = \langle \psi|\rho_2|\psi\rangle$.
Hence for any two pure states
their fidelity is equal to  their squared overlap,
   $F(\psi, \phi) =|\langle \psi|\phi \rangle |^2=:\kappa$.

Fidelity enjoys several important properties \cite{Uh76,inf1,inf2,Uh83a,Jo94}, which
can also be proved on state spaces of unital C$^*$-algebras.
Some of them are:
\newcounter{fidelprop}   
\begin{list}{\bf \roman{fidelprop})}{\usecounter{fidelprop}}
\item
 {\bf Bounds:} $0\le F(\rho_1,\rho_2)\le 1$.  Furthermore  $F(\rho_1,\rho_2)=1$
iff $\rho_1=\rho_2$,
while $F(\rho_1,\rho_2)=0$
iff $ \mbox{supp}(\rho_1) \perp \mbox{supp}(\rho_2)$.
\item
 {\bf Symmetry:} $F(\rho_1,\rho_2)= F(\rho_2,\rho_1)$.

\item
  {\bf Unitary invariance:}
        $F(\rho_1,\rho_2)=F(U\rho_1U^{\dagger},U\rho_2U^{\dagger})$, for any unitary operator $U$.

\item
 {\bf Concavity:}  $F(\rho,a\rho_1+(1-a)\rho_2)\ge
             a F(\rho,\rho_1)+(1-a)F(\rho,\rho_2)$, for $a \in [0,1]$.

\item
  {\bf Multiplicativity:}
          $F(\rho_1 \otimes \rho_2,\rho_3 \otimes \rho_4)=
          F(\rho_1,\rho_3) F(\rho_2,\rho_4)$.

\item
{\bf Joint concavity:}
     $\sqrt{F}(a\rho_1+(1-a)\rho_2, a\rho'_1+(1-a)\rho'_2) \geq
       a \sqrt{F}(\rho_1,\rho'_1)+(1-a)\sqrt{F}(\rho_2,\rho'_2)$, for $a \in [0,1]$.

 \end{list}
For further analysis of fidelity properties it is instructive
to work with eigenvalues of a
matrix $\sqrt{\rho_1^{1/2}\rho_2  \rho_1^{1/2}}$.
Let us denote them by $\lambda_i, \  i=1,\dots,N$.
This matrix is positive so its eigenvalues and singular
values coincide.
Unless otherwise stated, we tacitly assume that $\lambda_1 \geq \lambda_2 \geq
\dots \geq \lambda_N$.
The root fidelity  reads
\begin{equation}
\label{rootfid}
 \sqrt {F(\rho_1,\rho_2)}   =   \tr \sqrt{\sqrt{\rho_1}
\rho_2\sqrt{\rho_1}}
   =  \sum_{i=1}^N \lambda_i .
\end{equation}
Squaring this equation one obtains a compact expression for fidelity,
\begin{equation}
\label{eqn:fidelity-sum-singular}
  F(\rho_1,\rho_2)    =  \left( \sum_{i=1}^N \lambda_i \right)^2
 =  \tr \rho_1\rho_2 + 2 \sum_{i<j} \lambda_i \lambda_j ,
\end{equation}
where we have taken into account that
$\tr \rho_1\rho_2 = \tr \sqrt{\rho_1}\rho_2\sqrt{\rho_1} = \sum_{i=1}^N
\lambda_i^2$.
The matrix  $\sqrt{\rho_1}\rho_2\sqrt{\rho_1}$ is similar to
 $\rho_1\rho_2$ and they share the same set of $N$ eigenvalues.

\section{Bounds for fidelity}\label{sec:bounds}
We shall need some further algebraic definitions. For any matrix $X$ of size
$N$
with a set of eigenvalues $\{ \lambda_1, \dots, \lambda_N \}$
we define elementary symmetric functions $s_m(X)$ as the elementary symmetric function of its eigenvalues
\cite[Def.~1.2.9]{hj}.
For instance, the second and third elementary symmetric functions read
\begin{eqnarray}
\label{symmfun}
  s_2(X) &=& \sum_{i<j} \lambda_i \lambda_j , \label{s_2}\\
  s_3(X) &=& \sum_{i<j<k} \lambda_i \lambda_j \lambda_k .
    \label{s33}
\end{eqnarray}
For any matrix of rank $r$ the highest non-vanishing symmetric function reads $s_r(X) =
\prod_{i=1}^r \lambda_i$.
In the generic case $r = N$ we have $s_N(X) = \det(X)$.

In this section we shall list several bounds for fidelity,
some of which are well known in the literature.
Let us start by stating a simple result,
\begin{equation}
    F(\rho_1,\rho_2) \le \tr \rho_1 \tr \rho_2 ,
  \end{equation}
which follows directly from Fact \ref{fact:2} (see Appendix A)
if we set $\nu=1/2$. This fact implies
the property $ F(\rho_1,\rho_2) \le 1$.

Expression (\ref{eqn:fidelity-sum-singular}) implies
the  following lower bound
\begin{equation}
  \tr \rho_1 \rho_2 \leq F(\rho_1,\rho_2)  
  \leq N \tr|\rho_1 \rho_2|.
  \label{eqn:bound-tr-prod}
\end{equation}
To get the upper bound we use Fact \ref{fact:3}  (see Appendix A)
and set $\nu=1/2$ to obtain
$||\sqrt{\rho_1} \sqrt{\rho_2}||_1^2  \le    N||\rho_1 \rho_2||_1$.

Let us now denote the spectra of  the states $\rho_1$ and $\rho_2$,
by  vectors $\vec p$ and $\vec q$, respectively.
The fidelity between them is then bounded by
the classical fidelity between diagonal density matrices \cite{MMPZ08}
\begin{equation}
 F ( p^{\uparrow}, q^{\downarrow})    \le  F(\rho_1,\rho_2) \leq
 F ( p^{\uparrow}, q^{\uparrow}) ,
\label{fidboth}
\end{equation}
where the arrows up (down)
indicate that the eigenvalues are put in the nondecreasing (nonincreasing)
order.

The lower bound in (\ref{eqn:bound-tr-prod}) can be improved,
since the following result is true \cite{Uhlm00}
\begin{equation}
    F(\rho_1,\rho_2) \ge \tr \rho_1 \rho_2
+ \sqrt{2} \sqrt { (\tr \rho_1\rho_2)^2- \tr \rho_1\rho_2\rho_1\rho_2}  .
\label{lowerbis}
  \end{equation}
The above inequality is saturated for any pair
of one--qubit states. Furthermore, the above inequality is an
equality if the rank of $\rho_1 \rho_2$ does not exceed two.
On the other hand, the inequality is strict if  that rank is larger
than two --- see Appendix E.
For completeness we present the simple proof
of inequality (\ref{lowerbis}) in Appendix B.

Another lower  bound is obtained if the
rank of $\rho_1 \rho_2$ is exactly $r$. If $s_r$ denotes
the $r^{\text{th}}$ elementary symmetric function then
\medskip
\begin{equation}
\label{determin}
F(\rho_1,\rho_2)  \geq
\tr \rho_1 \rho_2 + r(r-1) \sqrt[r]{s_r(\rho_1 \rho_2)} .
\end{equation}
This bound is proved in Appendix C.
If both states are generic, i.e. if they are of the maximal rank
the above formula reads
\begin{equation}
F(\rho_1,\rho_2)  \geq
\tr \rho_1 \rho_2 + N(N-1) \sqrt[N]{\det \rho_1 \det \rho_2}.
\end{equation}

The key result of this paper consist in the following upper bound,
in a sense complementary to (\ref{lowerbis}).

\begin{theorem}\label{prop:main-theorem}
  For any density matrices $\rho_1$ and $\rho_2$ we have
  \begin{equation}
    F(\rho_1,\rho_2) \leq \
\tr \rho_1 \rho_2 + \sqrt{(1-\tr \rho_1^2)(1-\tr \rho_2^2)}.
    \label{eqn:main-theorem}
  \end{equation}
\end{theorem}

Before presenting the proof in the subsequent section
let us first note that the bound is saturated if at least one of the states
is pure. Furthermore, an equality holds for any two mixed states
of size $N=2$. To show this property  observe that in this
case the sum in (\ref{eqn:fidelity-sum-singular})
consists of a single term $2\lambda_1\lambda_2=2\sqrt{{\rm det}(\rho_1
\rho_2)}
= \sqrt{2{\rm det}(\rho_1)} \sqrt{2{\rm det}(\rho_2)}$.
Since for any one-qubit state one has $2{\rm det}(\rho)=1-\tr\rho^2$
an equality in (\ref{eqn:main-theorem}) follows.
This fact was already known to H{\"u}bner \cite{Hu92}.
In a similar way we treat the more general case of $N = 3$
in Appendix \ref{sec:app3}, for which
some other equations for fidelity are derived.

\section{Proof of the main upper bound}


The notion of the second symmetric function (\ref{s_2})
allows us to write the expression
\begin{equation}
 [(\tr X)^2 - (\tr X^2)] = 2 s_2(X) .
    \label{sym2X}
  \end{equation}
Note that if $X$ has nonnegative eigenvalues then $s_2(X) \geq 0$.

Using (\ref{sym2X}) we can rewrite fidelity
\begin{equation}
F(\rho_1, \rho_2) = \tr \rho_1 \rho_2 + 2 s_2 \left(\sqrt{\rho_1^{1/2} \rho_2
\rho_1^{1/2}} \right),
\label{Fs1}
\end{equation}
and
\begin{equation}
\sqrt{(1-\tr \rho_1^2)(1-\tr \rho_2^2)}
= 2 \sqrt{s_2(\rho_1) s_2(\rho_2)}.
\label{Fs2}
\end{equation}
Thus the Theorem \ref{prop:main-theorem} can be equally expressed as an
inequality
\begin{equation}\label{main-as-ineq}
s_2 \left(\sqrt{\rho_1^{1/2} \rho_2 \rho_1^{1/2}} \right) \leq
\sqrt{s_2(\rho_1) s_2(\rho_2)}.
\end{equation}

The proof of (\ref{main-as-ineq}) is decomposed into two Lemmas, the proof of
which can be found in Appendix \ref{sec:LemmaProof}.

\begin{lemma}\label{lemma:s2(X)<s2(d(A)d(B))}
For given density matrices $\rho_1, \rho_2$ with eigenvalues $p_1, \dots , p_n$
and  $q_1, \dots , q_n$ respectively
\begin{equation}
s_2 \left(\sqrt{\rho_1^{1/2}\rho_2 \rho_1^{1/2}} \right) \leq  s_2
\left(\sqrt{\diag(p) \diag(q)} \right),
\end{equation}
where $\diag(p)$ and $\diag(q)$ denote diagonal matrices with entries on diagonal
$p_1, \dots , p_n$ and  $q_1, \dots , q_n$ respectively.
\end{lemma}
\begin{lemma}\label{lemma:s2<sqrt(s2)}
With notation as in Lemma \ref{lemma:s2(X)<s2(d(A)d(B))}, we have
\begin{equation}
s_2 \left(\sqrt{\diag(p) \diag(q)} \right) \leq \sqrt{s_2(\diag(p))
s_2(\diag(q))} =  \sqrt{s_2(\rho_1 ) s_2(\rho_2)}.
\end{equation}
\end{lemma}
\bigskip

\proofof{Theorem \ref{prop:main-theorem}}{
For given density matrices $\rho_1$ and $\rho_2$ with eigenvalues $p_1, \dots , p_n$
and  $q_1, \dots , q_n$ respectively.
We denote diagonal matrices with entries on diagonal $p_1, \dots , p_n$ and
$q_1, \dots , q_n$ as $\diag(p)$ and $\diag(q)$ respectively.
\begin{eqnarray*}
F(\rho_1, \rho_2) &=& \tr \rho_1 \rho_2 + 2 s_2 \left(\sqrt{\rho_1^{1/2} \rho_2
\rho_1^{1/2}} \right) \\
&\leq& \tr \rho_1 \rho_2 + 2 s_2 \left(\sqrt{\diag(p) \diag(q)} \right) \\
&\leq& \tr \rho_1 \rho_2 + 2 \sqrt{s_2(\rho_1 ) s_2(\rho_2)} = \tr \rho_1
\rho_2 + \sqrt{(1-\tr \rho_1^2)(1-\tr \rho_2^2)}.
\end{eqnarray*}
Making use of Lemma \ref{lemma:s2(X)<s2(d(A)d(B))} for the first inequality
and  of Lemma \ref{lemma:s2<sqrt(s2)} for the second one
we arrive at the inequality (\ref{eqn:main-theorem}).
}

\section{Sub-- and super--fidelity and their properties}

\subsection{Definition and basic facts}

We shall start this section with a general {\bf definition}.
For any two hermitian operators $A$ and $B$
let us define two quantities
\begin{eqnarray}
  E(A,B) &=& \tr AB + \sqrt{2[(\tr AB)^2  - \tr ABAB]} ,
  \label{eqn:sub-fide-def} \\
  G(A,B) &=& \tr AB + \sqrt{(\tr A)^2-\tr A^2}  \sqrt{(\tr B)^2-\tr B^2} .
  \label{eqn:quasi-fide-def}
\end{eqnarray}

For any two density operators their traces are equal to unity,
so  $E(\rho_1,\rho_2)$ and
 $G(\rho_1,\rho_2) $ have lower bound
(\ref{lowerbis}) and upper bound  (\ref{eqn:main-theorem}),
respectively. Thus both universal bounds for fidelity can be rewritten as
\begin{equation}
E(\rho_1,\rho_2) \le  F(\rho_1 , \rho_2 )  \le   G(\rho_1 , \rho_2 ) .
  \label{quasi-ineq}
\end{equation}
Note that both bounds require the evaluation of three traces only,
so they are easier to compute than the original fidelity.
As shown in Section \ref{sec:bounds}
for $N=2$ all three quantities are equal, so we propose to call
$E(\rho_1 , \rho_2 )$ and
$G(\rho_1 , \rho_2 )$ as {\sl sub--} and {\sl super--fidelity}.
These names are additionally motivated by the
following appealing properties:

\newcounter{superfidelprop}
\begin{list}{\bf \roman{superfidelprop}')} {\usecounter{superfidelprop}}
\item
{\bf Bounds:} $0 \le E(\rho_1,\rho_2) \le 1$  and
                    $0 \le G(\rho_1,\rho_2) \le 1$.

\item
    {\bf Symmetry:} $E(\rho_1,\rho_2)= E(\rho_2,\rho_1)$  and
$G(\rho_1,\rho_2)= G(\rho_2,\rho_1)$.

\item
    {\bf Unitary invariance:}
$E(\rho_1,\rho_2)=E(U\rho_1U^{\dagger},U\rho_2U^{\dagger})$
and  $G(\rho_1,\rho_2)=G(U\rho_1U^{\dagger},U\rho_2U^{\dagger})$, for any unitary operator $U$.

\item
    {\bf Concavity:}
\begin{prop}\label{prop:qfid-concave} Sub-- and super--fidelity are  concave,
that is for $A,B,C,D \in \states_N$ and $\alpha \in [0,1]$ we have 
\begin{eqnarray}
  E(A,\alpha B + (1-\alpha)C) &\geq& \alpha E(A,B) + (1-\alpha) E(A,C),
  \label{eqn:subfid-conca-prop} \\
  G(A,\alpha B + (1-\alpha)C) &\geq& \alpha G(A,B) + (1-\alpha) G(A,C).
  \label{eqn:qfid-conca-prop}
\end{eqnarray}
\end{prop}

\item
    {\bf Properties of the tensor product:}
\begin{prop}\label{prop:qfid-smul} Super--fidelity is super--multiplicative, that is for $A,B,C,D \in \states_N$
\begin{equation}
G(A \otimes B, C \otimes D) \geq G(A,C) G(B,D),
\end{equation}
\end{prop}
while
\begin{prop}\label{prop:subfid-sub} Sub--fidelity is sub--multiplicative, that is for $A,B,C,D \in \states_N$
\begin{equation}
E(A \otimes B, C \otimes D) \leq E(A,C) E(B,D).
\end{equation}
\end{prop}

\end{list}
Properties {\bf i')}, {\bf ii')} and {\bf iii')} follow from the properties of $\tr AB$ and definitions
(\ref{eqn:sub-fide-def}) and (\ref{eqn:quasi-fide-def}).
In this section we prove properties {\bf iv')} and {\bf v')}.

\proofof{Proposition \ref{prop:qfid-concave}}{
The definitions (\ref{eqn:sub-fide-def}) and (\ref{eqn:quasi-fide-def}) can be
rewritten in terms of the aforementioned elementary symmetric functions
(\ref{symmfun})
using relation (\ref{sym2X}),
\begin{eqnarray}
  E(A,B) &=& \tr AB +  2 \sqrt{ s_2(AB)} , \label{sub-s2} \\
  G(A,B) &=& \tr AB + 2 \sqrt{s_2(A) s_2(B)} .
\label{quasi-s2}
\end{eqnarray}
In general $r^{\text{th}}$ root of $s_r$ is concave on the cone
of  positive operators \cite{Marcus}.
This implies concavity of $G$ directly. To get concavity of $E$
we can replace matrix $AB$ by the similar matrix $A^{1/2} B A^{1/2}$
which is positive. Using the concavity of $\sqrt{s_2(A^{1/2} B A^{1/2})}$ we obtain the result.
}

\proofof{Proposition \ref{prop:qfid-smul}}{
First we note that super--fidelity is not multiplicative. As an example we can
take
\begin{equation}
A =
\left(\begin{array}{c c}
1 & 0\\
0 & 0\\
\end{array}
\right),\
B =
\left(\begin{array}{c c}
\frac{1}{2} & 0\\
0 & \frac{1}{2}\\
\end{array}
\right),\
C =
\left(\begin{array}{c c}
0 & 0\\
0 & 1\\
\end{array}
\right),\
D =
\left(\begin{array}{c c}
\frac{1}{2} & 0\\
0 & \frac{1}{2}\\
\end{array}
\right),
\end{equation}
in which case we have
\begin{equation}
\frac{1}{2} = G(A \otimes B, C \otimes D) > G(A,C) G(B,D) = 0.
\end{equation}
To prove the proposition we write
\begin{equation}
G(A \otimes B, C \otimes D) =
 \tr AC  \tr BD + \sqrt{(1-\tr A^2  \tr B^2)(1-\tr C^2  \tr D^2)},
\end{equation}
and
\begin{equation}
G(A,C) G(B,D)
= \left(\tr AC + \sqrt{(1-\tr A^2 )(1-\tr C^2)} \right)
\left(\tr BD + \sqrt{(1-\tr B^2 )(1-\tr D^2)} \right).
\end{equation}
Denoting
$\tr A^2 = \alpha, \tr B^2 = \beta, \tr C^2 = \gamma$ and $\tr D^2 = \delta$
we have to show that
\begin{eqnarray*}
\sqrt{(1-\alpha \beta)(1-\gamma \delta)}
&\geq&
\tr AC \sqrt{(1-\beta )(1-\delta)}
 + \tr BD \sqrt{(1-\alpha)(1-\gamma)} \nonumber \\
&& + \sqrt{(1-\alpha )(1-\gamma)(1-\beta )(1-\delta)}.
\end{eqnarray*}
Now from Fact \ref{fact:holder} with $a = 2$ (see Appendix B)
one has
\begin{equation}
\tr AC \leq  \sqrt{\tr A^2   \tr C^2} = \sqrt{\alpha \gamma}
\end{equation}
and
\begin{equation}
\tr BD \leq  \sqrt{\tr B^2   \tr D^2} = \sqrt{\beta \delta}.
\end{equation}
Thus it is enough to show that
\begin{eqnarray}\label{eqn:mainIneq}
\sqrt{(1-\alpha \beta)(1-\gamma \delta)} 
&\geq&
\sqrt{\alpha \gamma} \sqrt{(1-\beta )(1-\delta)}
+\sqrt{\beta \delta} \sqrt{(1-\alpha)(1-\gamma)} \nonumber \\
&& + \sqrt{(1-\alpha )(1-\gamma)(1-\beta )(1-\delta)}.
\end{eqnarray}
We define two vectors
\begin{equation}
X =
\left(\begin{array}{c}
 \sqrt{\alpha}\sqrt{1-\beta} \\
 \sqrt{\beta} \sqrt{1 - \alpha}\\
 \sqrt{1-\alpha}\sqrt{1-\beta}
 \end{array}
\right)
\text{ and }
Y =
\left(\begin{array}{c}
 \sqrt{\gamma}\sqrt{1-\delta} \\
 \sqrt{\delta} \sqrt{1 - \gamma}\\
 \sqrt{1-\gamma}\sqrt{1-\delta}
 \end{array}
\right).
\end{equation}
Note that
\begin{equation}\label{eqn:xy}
\scalar{X}{Y} = \sqrt{\alpha \gamma} \sqrt{(1-\beta )(1-\delta)}
+\sqrt{\beta \delta} \sqrt{(1-\alpha)(1-\gamma)}
+\sqrt{(1-\alpha )(1-\gamma)(1-\beta )(1-\delta)}
\end{equation}
and
\begin{equation} \label{eqn:xx}
\scalar{X}{X} = (1-\alpha \beta) \text{ and } \scalar{Y}{Y} = (1-\gamma
\delta).
\end{equation}
Now by combining (\ref{eqn:xx}) with (\ref{eqn:xy}) and using Cauchy--Schwarz
inequality
\begin{equation}
\sqrt{\scalar{X}{X} \scalar{Y}{Y}} \geq \scalar{X}{Y},
\end{equation}
we obtain (\ref{eqn:mainIneq}).
}

\proofof{Proposition \ref{prop:subfid-sub}}{
To show sub--multiplicativity of sub--fidelity
we write the definition (\ref{eqn:sub-fide-def})
for a tensor product,
\begin{eqnarray*}
E(A\otimes B, C\otimes D) &=& \tr [(A\otimes B) (C\otimes D)]
\\ 
&&
+ \sqrt{2[ \bigl( \tr (A\otimes B)(C\otimes D)\bigr)^2  -
\tr (A\otimes B)(C\otimes D)( A\otimes B)(C\otimes D)]} 
\\
&=&
\tr AC \tr BD + \sqrt{2[(\tr AC \tr BD)^2  - \tr ACAC \tr BDBD]} .
\end{eqnarray*}
The product of two sub--fidelities reads
\begin{eqnarray*}
\lefteqn{E(A,C)  E(B,D) = (\tr AC + \sqrt{2[(\tr AC)^2  - \tr ACAC]} )(\tr BD +
\sqrt{2[(\tr BD)^2  - \tr BDBD]} ) }\\
&=&
\tr AC\tr BD  + \tr AC \sqrt{2[(\tr BD)^2  - \tr BDBD]} + \tr BD \sqrt{2[(\tr
AC)^2  - \tr ACAC]}  \\
&& + \sqrt{2[(\tr AC)^2  - \tr ACAC]}\sqrt{2[(\tr BD)^2  - \tr BDBD]}.
\end{eqnarray*}
For short we denote
\begin{displaymath}
\begin{array}{lll}
\alpha  =   \tr AC,  & a   =  \tr ACAC, \\ 
\beta   =   \tr BD,  & b   =  \tr BDBD.
\end{array}
\end{displaymath}

We have $\alpha^2 \geq a$ and $\beta^2 \geq b$.
By rewriting above expressions in the new notation we obtain
\begin{eqnarray*}
E(A\otimes B, C\otimes D) &=&
\alpha \beta + \sqrt{2[\alpha^2 \beta^2  - a b ]}
\end{eqnarray*}
and
\begin{eqnarray*}
E(A,C) E(B,D) &=&
\alpha \beta  + \alpha  \sqrt{2[\beta^2  - b]} + \beta \sqrt{2[\alpha ^2  - a]}
+ \sqrt{2[\alpha ^2  - a]}\sqrt{2[\beta^2  - b]}.
\end{eqnarray*}
Now we write
\begin{eqnarray*}
\sqrt{2[\alpha^2 \beta^2  - a b ]} &=& \sqrt{2[\alpha^2 \beta^2 - ab + ab - ab
- \alpha^2 b + \alpha^2 b - \beta^2 a + \beta^2 a ]} \\
&=& \sqrt{2[(\alpha^2-a)(\beta^2-b) + b(\alpha^2-a)   + a (\beta^2 -b) ]}.
\end{eqnarray*}
Making use of subadditivity of square root we obtain
\begin{eqnarray*}
\sqrt{2[\alpha^2 \beta^2  - a b ]}
&\leq& \sqrt{2(\alpha^2-a)(\beta^2-b)} + \sqrt{2 b(\alpha^2-a)}
  + \sqrt{2 a (\beta^2 -b) ]} .
\end{eqnarray*}
Because $2<4$, $a \leq \alpha^2$ and $b \leq \beta^2$ we get
\begin{eqnarray*}
\sqrt{2[\alpha^2 \beta^2  - a b ]}
&\leq& \sqrt{4(\alpha^2-a)(\beta^2-b)} + \beta\sqrt{2 (\alpha^2-a)}   +
\alpha\sqrt{2 (\beta^2 -b) ]}.
\end{eqnarray*}
And as a result we obtain the desired inequality
\begin{eqnarray*}
E(A\otimes B, C\otimes D) &\leq& E(A,C) E(B,D).
\end{eqnarray*}
}

For any pair of Hermitian operators $X_1$ and $X_2$
let us now define a  quadratic {\sl Lorentz form}
\begin{equation}
 (X_1,X_2)_L := [(\tr X_1)  (\tr X_2) - \tr X_1 X_2] .
    \label{quadform}
\end{equation}

To find out the motivation standing behind this name
let us expand a Hermitian operator $X$ in an operator basis,
$X=\sum_{j=0}^{N^2-1} a_j H_j$.
We assume that the basis is orthogonal, $\tr H_j H_k=\delta_{jk}$,
the first operator is proportional to identity,  $H_0={\1}/\sqrt{N}$,
 and all other operators $H_j$ are traceless.
Then the form (\ref{quadform}) gives
\begin{equation}
 (X,X)_L = (\tr X)^2 - \tr X^2  = \frac{(N-1)}{N} a_0^2 - \sum_{j=1}^{N^2-1} a_j^2 ,
    \label{quadform2}
  \end{equation}
which is of Minkowski--Lorentz type.
By the help of this notion, super--fidelity can be written as
\begin{equation}
  G(A,B) = \tr AB + [(A,A)_L (B,B)_L]^{1/2} ,
  \label{quasi-lor}
\end{equation}
while sub--fidelity reads
\begin{equation}
  E(A,B) = \tr AB + [2(AB,AB)_L]^{1/2} .
  \label{sub-lor}
\end{equation}

\medskip

The \emph{forward cone} with respect to the form
(\ref{quadform}) is given by operators $X$ satisfying
\begin{equation}
 (X,X)_L  \ge 0
\hbox{ and }
\tr X \ge 0 .
    \label{pos_cone}
  \end{equation}

Since the density matrices are normalized, $\tr \rho=1$,
the form $(\rho,\rho)_L$ is non--negative.

For a Lorentz form any two forward directed Hermitian matrices
$A$ and $B$ satisfy
\begin{equation}
 (A,A)_L  (B,B)_L  \le  [(A,B)_L]^2 .
  \end{equation}
Substituting this bound into expression (\ref{quasi-lor})
we arrive at an upper bound for super--fidelity
\begin{equation}
  G(A,B) \le  \tr AB  + (\tr A)(\tr B) - (\tr AB) =  (\tr A)(\tr B) .
  \label{Gbound}
\end{equation}
For the case of normalized density matrices, $\tr \rho=1$
we get $G(\rho_1,\rho_2)\le 1$.

Using the bound (\ref{determin}) for density operators we introduce
a third quantity
\begin{equation} \label{def:E'}
E'(A,B) = \tr AB + r(r-1)\sqrt[r]{s_r(A B)},
\end{equation}
where $r$ is the rank of matrix $AB$. Note that for $r=2$ this
formula is reduced to an expression (\ref{sub-s2}) for sub--fidelity,
 hence in this case $E'=E$.

Since $(s_r(X))^{1/r}$ is concave for density operators
we infer that the quantity  $E'(A,B)$,  defined in equation (\ref{def:E'}),
is separately concave in $A$ and in $B$.

\subsection{Examples and classical analogues}
\begin{figure}[htbp]
\centerline{\epsfig{file=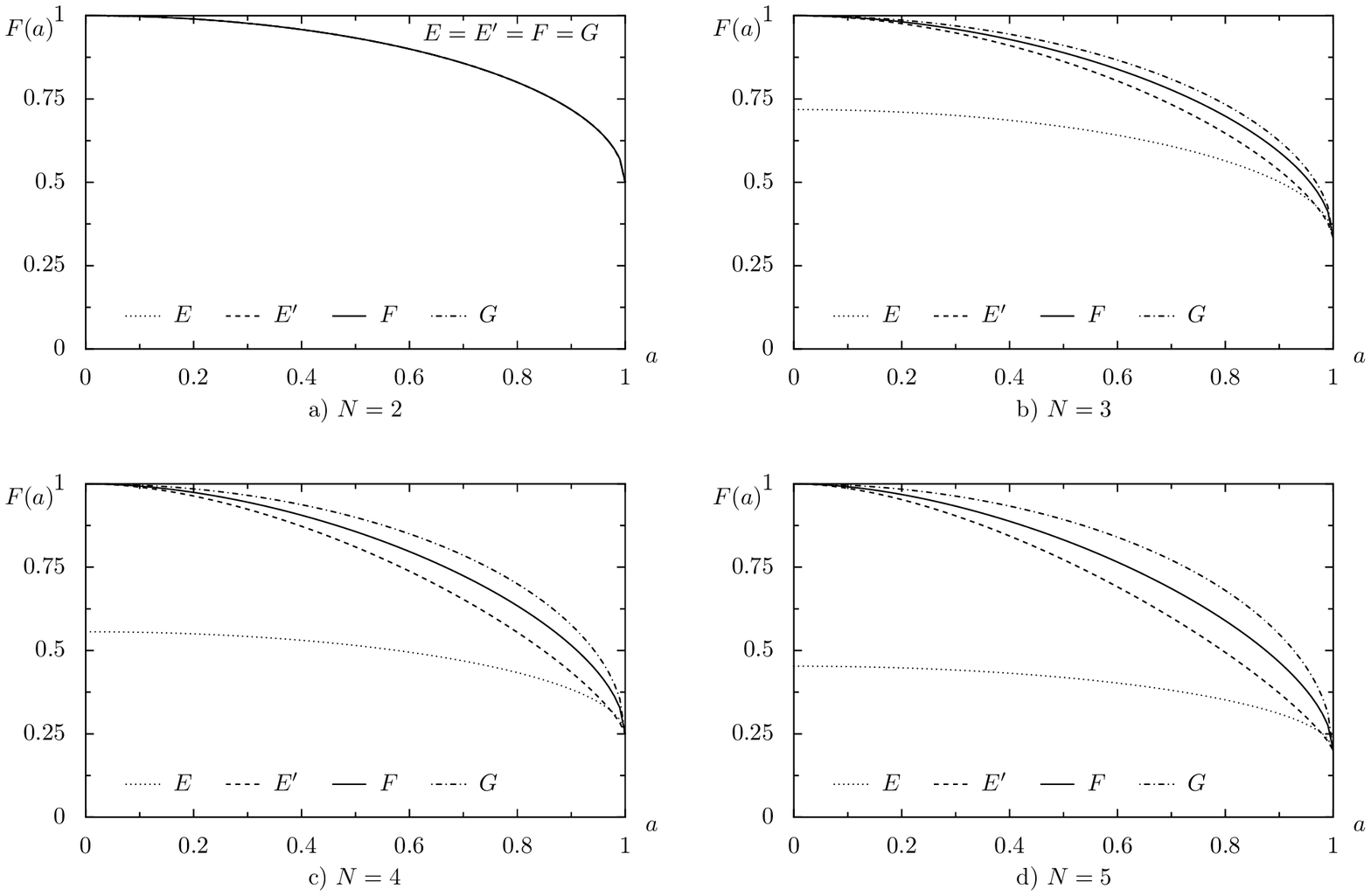, width=12cm}}
\vspace*{13pt}
\fcaption{The comparison of sub--fidelity $E$, bound $E'$, fidelity $F$ (solid
line)
   and  super--fidelity $G$.
Each plot shows these quantities calculated for the maximally mixed state and
a
state  (\ref{rhoa}) depending on the parameter $a$.
For a one--qubit system, case a)  $N=2$,
one has  $E=E'=F=G$.  Note the difference between these quantities
shown for  $N=3,4,5$.
In this case  $E>E'$  for $a$ close to unity.}
  \label{fig:ex-fid-compare}
\end{figure}

%
To observe sub-- and super--fidelity  in action
consider a family of mixed states
\begin{equation}
  \rho_a =a|\psi\rangle \langle \psi| + (1-a) {\mathbbm I}/N ,
  \label{rhoa}
\end{equation}
which interpolates between arbitrary pure state $|\psi\rangle$
and the maximally mixed state.
It is straightforward to compute the fidelity
between the state $\rho_a$ and the maximally mixed state
$\rho_*:={\mathbbm I}/N$,
\begin{equation}
  F(\rho_a,\rho_*)  =   \frac{1}{N^2} \left( \sqrt{(N-1)a + 1} +
(N-1)\sqrt{1-a} \right)^2 ,
  \label{Fa}
\end{equation}
as well as other bounds
\begin{eqnarray}
E(\rho_a,\rho_*)
&=&
 \frac{1}{N} + \sqrt{2} \frac{1}{N} \sqrt{1 - \frac{1}{N}} \sqrt{1 - a^2}  ,  \\
E'(\rho_a,\rho_*)
&=&
 \frac{1}{N}  + \left(1 - \frac{1}{N}\right) \sqrt[N]{((N-1)a + 1)(1-a)^{N-1} } , \\
G(\rho_a,\rho_*)
&=&
\frac{1}{N} + \left(1 - \frac{1}{N}\right) \sqrt{1 - a^2} .
\label{Ga}
\end{eqnarray}
These results are plotted in Fig. \ref{fig:ex-fid-compare} for $N=2,3,4,5$.

For $N=2$ all these quantities coincide, and the quality of the approximation
goes down with the system size $N$, as expected.
Looking at the graph one could imagine that relation $E\le E'$
is fulfilled. However, such an equality does not hold
as we found a counter example: the pair of states analyzed in the figure
with parameter $a$ very close to unity.

One may work out several other examples, for which
sub-- and super--fidelity are easy to find. Explicit formulas are
simple in the case of two commuting density matrices $\rho_p$
and $\rho_q$
with spectra given by vectors $\vec p$ and $\vec q$, respectively.
In such a classical case these quantities read
\begin{eqnarray}
E(\rho_p, \rho_q)
& = &
 \sum_{i=1}^N p_i q_i +\sqrt{2 \left[ \left(
 \sum_{i=1}^{N} p_i q_i  \right)^2-
 \sum_{i=1}^N p_i^2 q_i ^2 \right] } , \\
F(\rho_p,\rho_q)
&=&
 \left(  \sum_{i=1}^N \sqrt{p_i q_i} \right)^2 ,\\
 G(\rho_p,\rho_q)
&=&
 \sum_{i=1}^N p_i q_i +
 \sqrt{\left(1- \sum_{i=1}^N p_i^2\right) \left(1- \sum_{i=1}^N q_i^2\right)}.
 \label{classical}
\end{eqnarray}

\subsection{The difference $G-F$}

In view of the inequality (\ref{quasi-ineq})
it is natural to ask how big the difference $G-F$ might be.
Since both quantities coincide if one of the states is pure,
let us analyze the case of two mixed states living
in orthogonal subspaces.

More precisely,  let us fix an even dimensionality of the Hilbert space
$N=2M$,
and define two diagonal states, each supported in $M$
dimensional space,
$\rho_1=\frac{2}{N}{\rm diag}(1,\dots,1,0,\dots,0)$
and
$\rho_2=\frac{2}{N}{\rm diag}(0,\dots,0,1,\dots,1)$.
Since they are supported by orthogonal subspaces
their fidelity vanishes, $F(\rho_1,\rho_2)=0$.
On the other hand the definition (\ref{eqn:quasi-fide-def})
gives their super--fidelity
\begin{equation}
G(\rho_1,\rho_2) = \frac{N-2}{N},
\end{equation}
equal in this case to the difference $G-F$.
As expected for $N=2$ we get $G=F=0$. However,
for $N$ large enough the difference $G-F$
may become arbitrarily close to unity.

Thus working with super--fidelity $G$ in place of fidelity  $F$
one needs to remember that this approximation works fine for 
small systems or where at least one of the states is pure enough.

\section{On measurement methods}

\subsection{Associated physical observables}

Here we shall  shortly discuss possibilities of measurement of 
both  sub-- and super--fidelities in physical experiments.
The approach below follows the techniques used in state spectrum estimation \cite{EkertEtAl} and  
nonlinear entanglement detection and/or estimation which has been developed
significantly last years (see \cite{ADH-Single-Witness,ASH-Generalised-Entropies} and references therein). 
Those approaches exploited the properties of SWAP operator and other permutation unitary 
operations to get the properties of single state rather than the relation of different states. 
There were little exceptions: one was a quantum network measurement of an overlap of the two 
states \cite{EkertEtAl}. Here we shall follow the latter idea since we 
want to estimate  the {\it distance} of two different quantum states.
In particular we shall see that it is  possible to measure these quantities
with help of not more than two collective observables. 
This fact may be helpful in experimental comparison of two different stationary 
sources of quantum states. Quite remarkably, as we shall see below, with help of similar techniques,
super--fidelity can be represented by only three  
experimental probabilities which makes it very friendly from 
an experimental point of view.

We start by providing a simple example. First one can see that to calculate
sub-- and super--fidelity it is necessary to calculate the values of the terms of
the form $\tr AB$. Let $A,B\in M_2(\Cplx)$. In this case
\begin{equation}
A=\left(
\begin{array}{ll}
 a_{11} & a_{12} \\
 a_{21} & a_{22}
\end{array}
\right)
,
B=\left(
\begin{array}{ll}
 b_{11} & b_{12} \\
 b_{21} & b_{22}
\end{array}
\right)
\end{equation}
and
\begin{equation}
 \tr AB = \tr \left[\left(
\begin{array}{ll}
 a_{11} b_{11}+a_{12} b_{21} & a_{11} b_{12}+a_{12} b_{22} \\
 a_{21} b_{11}+a_{22} b_{21} & a_{21} b_{12}+a_{22} b_{22}
\end{array}
\right)\right] = a_{11} b_{11}+a_{21} b_{12}+a_{12} b_{21}+a_{22} b_{22}.
\end{equation}
On the other hand this value can be calculated using SWAP gate as
\begin{eqnarray}
 \tr \left[\text{SWAP} (A\otimes B) \right]&=&
\tr \left[\left(
\begin{array}{llll}
 1 & 0 & 0 & 0 \\
 0 & 0 & 1 & 0 \\
 0 & 1 & 0 & 0 \\
 0 & 0 & 0 & 1
\end{array}
\right) \left(
\begin{array}{llll}
 a_{11} b_{11} & a_{11} b_{12} & a_{12} b_{11} & a_{12} b_{12} \\
 a_{11} b_{21} & a_{11} b_{22} & a_{12} b_{21} & a_{12} b_{22} \\
 a_{21} b_{11} & a_{21} b_{12} & a_{22} b_{11} & a_{22} b_{12} \\
 a_{21} b_{21} & a_{21} b_{22} & a_{22} b_{21} & a_{22} b_{22}
\end{array}
\right)\right] \\
&=& \tr\left[
\left(
\begin{array}{llll}
 a_{11} b_{11} & a_{11} b_{12} & a_{12} b_{11} & a_{12} b_{12} \\
 a_{21} b_{11} & a_{21} b_{12} & a_{22} b_{11} & a_{22} b_{12} \\
 a_{11} b_{21} & a_{11} b_{22} & a_{12} b_{21} & a_{12} b_{22} \\
 a_{21} b_{21} & a_{21} b_{22} & a_{22} b_{21} & a_{22} b_{22}
\end{array}
\right)
\right]
 = \tr AB.
\label{SWAPproperty}
\end{eqnarray}
To address the question of measurability of the 
quantities (\ref{eqn:sub-fide-def}), (\ref{eqn:quasi-fide-def})
let us first recall the corresponding permutation operators
which we shall need subsequently.
The first one will be just SWAP operator (example of which is the SWAP gate presented above)
$V_{12}: {\cal H}_N \otimes 
{\cal H}_{N} \rightarrow {\cal H}_N \otimes {\cal H}_{N}$
which is defined by the action
\begin{equation}
V_{12}|\phi_{1}\rangle \otimes |\psi_{2}\rangle = |\psi_{2}\rangle \otimes
|\phi_{1}\rangle .
\label{swap}
\end{equation}
This is a Hermitian operator and as a such it represents an observable.
It  has a simple eigendecomposition in the form
\begin{equation}
V_{12}=P^{(+)}_{12} - P^{(-)}_{12} ,
\label{swap-eigendecomposition}
\end{equation}
where projections $P^{(\pm)}_{12}$ onto symmetric and antisymmetric
subspaces of ${\cal H}_N \otimes {\cal H}_{N}$ are
\begin{equation}
P_{12}^{\pm}=\frac{1}{2}({\mathbbm I}_{12} \pm V_{12}) . 
\label{sym-antisym-proj}
\end{equation}
Below we shall omit the indices and use the notation $V$ and $P^{(\pm)}$,
if it does  not lead to confusion.
An important property usually exploited in case of entanglement detection is that 
the formula (\ref{SWAPproperty}) holds for the 
SWAP operator $V$ of any dimension \cite{Werner}.
Apart form that operation we will also need a family of unitary permutation matrices 
$V_{1234}^{\pi}:{\cal H}_N^{\otimes 4} \rightarrow {\cal H}_{N}^{\otimes 4}$,
\begin{equation}
V_{1234}^{\pi} |\psi_{1}\rangle \otimes |\psi_{2} \rangle \otimes 
|\psi_{3} \rangle \otimes |\psi_{4}\rangle = 
|\psi_{\pi(1)}\rangle \otimes |\psi_{\pi(2)} \rangle \otimes 
|\psi_{\pi(3)} \rangle \otimes |\psi_{\pi(4)}\rangle , 
\end{equation}
where $\pi$ represents any chosen permutation of the indices $(1,2,3,4)$. 
For simplicity  we shall drop the indices 
using the notation $V^{\pi}$.

Let us define the set ${\cal S}$ of all eight permutations that do not map the sequence $(1,2,3,4)$
into a one having odd or even elements one after another.
For instance, the permutations defined by the ranges (2341) or (3214) belong to ${\cal S}$, 
while (2314) or (1423) do not. 
For a fixed  set ${\cal S}'\subset {\cal S}$ and some $\pi_{0} \in {\cal S}$ we define 
the following observable:
\begin{equation}
W^{{\cal S}',\pi_{0}}
=\frac{1}{2|{\cal S'}|} \left( \sum_{\pi \in {\cal S}'}V^{\pi}V^{\pi_{0}}V^{\pi} +
\sum_{\pi \in {\cal S}'}V^{\pi^{-1}}V^{\pi_{0}^{-1}}V^{\pi^{-1}} \right) .
\label{averaged-4-copy-obs}
\end{equation}
A special case is the observable 
$W^{\{\pi_{0}\},\pi_{0}}=(V^{\pi_{0}}+V^{\pi_{0}^{-1}})/2$ with $\pi_{0}$ being 
just some cyclic permutation (cf. \cite{ADH-Single-Witness} and references therein).
The choice of the permutation $\pi_{0}$ and/or the subset ${\cal S}'$ may be motivated by 
a specific physical situation. 

In the case of  single qubit sources ($N=2$) the observables (\ref{averaged-4-copy-obs}) have highly degenerated spectra and the corresponding eigenvectors have very symmetric forms. 
In particular the observable $W^{\{\pi_{0}\},\pi_{0}}$ has spectrum $\{1,-1,0 \}$ which means that its mean value requires probabilities of only two outcomes of incomplete 
von Neumann measurement.  The observable has the spectral decomposition 
$W^{\{\pi_{0}\},\pi_{0}}=Q^{(+)}-Q^{(-)}$ where support of the projector $Q^{(+)}$
is spanned by eigenvectors $\{ |\phi_{1}\rangle=|0000\rangle,|\phi_{2}\rangle=|1111\rangle,
|\phi_{3}\rangle=(|0111\rangle +|1011\rangle + |1101\rangle +|1110\rangle)/2,|\phi_{4}\rangle=
(|0011\rangle +|0110\rangle + |1001\rangle +|1100\rangle)/2,
|\phi_{5}\rangle=(|0101\rangle +|1010\rangle)/\sqrt{2},|\phi_{6}\rangle=\sigma_{x}^{\otimes 4}|\psi_{4}\rangle \}$
while the support of the second projector $Q^{(-)}$ (orthogonal to $Q^{(+)}$) corresponds to
$\{ I\otimes \sigma_{z}^{\otimes 2} \otimes I|\phi_{3}\rangle, I^{\otimes 2} \otimes \sigma_{z}^{\otimes 2}|\phi_{4}\rangle, I^{\otimes 3} \otimes \sigma_{z}|\phi_{5}\rangle, I\otimes \sigma_{z}^{\otimes 2} \otimes I|\phi_{6}\rangle \}$.

To illustrate how to measure the quantities $E$ and $G$ suppose now we can 
perform collective measurements on two and four copies 
of both quantum states. We plan  measurements that allow  two or four 
copies of analyzed  states to interact.
 Then (cf. \cite{EkertEtAl,ADH-Single-Witness,ASH-Generalised-Entropies} and references therein) 
the sub-- and super--fidelities can be 
represented in terms of averages of following observables, 
\begin{eqnarray}
  E(\rho_1,\rho_2) &=& \tr V \rho_1\otimes \rho_2 + \sqrt{2[(\tr V \rho_1 \otimes \rho_2)^2  - 
\tr W^{{\cal S},\pi_{0}}\rho_1 \otimes \rho_2\otimes \rho_1 \otimes \rho_2]} ,
  \label{eqn:sub-fide-exp} \\
  G(\rho_1,\rho_2) &=& \tr V \rho_1\otimes \rho_2 + \sqrt{1-\tr V \rho_1\otimes\rho_1}  \sqrt{1-\tr V \rho_2 \otimes \rho_2} .
  \label{eqn:quasi-fide-exp}
\end{eqnarray}

There are two simple but important observations to be made.
The sub--fidelity $E$ can be measured with help of two setups:
 (i) the one measuring the observable $V$ and (ii) the second one 
measuring observable $W^{{\cal S},\pi_{0}}$.
Each setup requires one source: setup (i) needs the source that creates, 
say, pairs $\rho_{1} \otimes \rho_2$,
while  setup (ii) requires a source  producing quadruples of the form, say, 
$\rho_{1} \otimes \rho_2 \otimes \rho_{1} \otimes \rho_{2}$.

Our scheme will work also for a worse source that produces
one of the  pairs (quadruples) $\{ \rho_{1} \otimes \rho_2, \rho_{2} \otimes \rho_1 \}$
 ($\{ \rho_{1} \otimes \rho_2 \otimes \rho_{1} \otimes \rho_{2},\rho_{2} \otimes \rho_1 \otimes 
\rho_{2} \otimes \rho_{1} \} $) at random  according to an unknown biased probability distribution,
which will not affect the results of the corresponding estimate for sub--fidelity.

The second observation is that the super--fidelity $G$ can  be measured with help
of {\it single} setup, namely the one that measures observable $V$, but requires
its application to three types of sources i.e. the ones creating pairs $\rho_{1}
\otimes \rho_1$, $\rho_{2} \otimes \rho_2$, and, say, $\rho_{1} \otimes
\rho_2$. Again, the last source may produce at random one of the pairs $\{
\rho_{1} \otimes \rho_2, \rho_{2} \otimes \rho_1 \}$ and this will not affect
the estimate for  super--fidelity.

It is very interesting to study the form of super--fidelity in terms
of directly measurable quantities, \emph{i.e.}  {\it probabilities}, since it has a
simple optical implementation. Let us introduce the probabilities of the
projection onto the antisymmetric subspace of ${\cal H}_N \otimes
{\cal H}_{N}$:
\begin{equation} 
p^{(-)}_{ij}=\tr P^{(-)} \rho_{i} \otimes \rho_{j},\ \  i, j = 1, 2.
\label{probability-antysymmetric}
\end{equation}
Then super--fidelity has a particularly nice form,
\begin{equation}
G(\rho_1,\rho_2) = 1-2\left( p^{(-)}_{12} - \sqrt{p^{(-)}_{11}p^{(-)}_{22}} \right) \ ,
\label{super-fidelity-3probabilities}
\end{equation} 
which is crucial for further discussion.
Note that the super--fidelity can be represented in terms of  only  {\it three}  probabilities
that can be measured in a single set-up. One can perform a simple 
consistency test by checking, whether the combination of
experimental probabilities satisfy  (up to error bars)  the condition
$p^{(-)}_{12} - \sqrt{p^{(-)}_{11}p^{(-)}_{22}}\leq 0.5$ --
otherwise one had an unphysical result,
 since super--fidelity can not be negative.
Note that the probability  $p^{(-)}_{11}$ has been already measured experimentally for two 
copies of composite systems in context of entanglement detection \cite{Bovino} or estimation \cite{Walborn}
under some assumptions about the nature of the sources. 
In subsection below we shall refer to the scheme analogous to the one utilized 
in Ref. \cite{Bovino}.

It is interesting to note that if the state $\varrho$ is of $d$-dimensional type, then 
reproduction of sub-- and super--fidelity {\it via} quantum tomography requires $2d^{2}-2$ independent quantities to be estimated since each of the two states is described by $d^{2}-1$ real parameters.
On the other hand, to find the quantities $E$ and $G$ in the way described above one requires only two or three independent real quantities (probabilities) to be estimated independly on how large
the dimension $d$ is. The price to be payed is, of course, that one must perform collective  
experiments. Preparation of reliabe setups of such experiments might be a good test for quantum 
engeneering.
   
\subsection{Measuring super--fidelity of states representing 
photons polarizations}

Consider now physical setup that would compare two states of 
polarization of single photon in terms of super--fidelity $G$.
In this case the density matrix is defined on Hilbert space 
isomorphic to $C^{2}$ where the horizontal (vertical) polarization, 
usually denoted by $|H \rangle$ ($|V \rangle$)
corresponds to the standard basis element $|0\rangle$ ($|1 \rangle$).
Suppose one has memoryless sources of two types $S_{i}$ ($i=1,2$)  sending 
photons in polarization states $\rho_{1}$, $\rho_{2}$ respectively.

\begin{figure}[ht]
\centering
\includegraphics{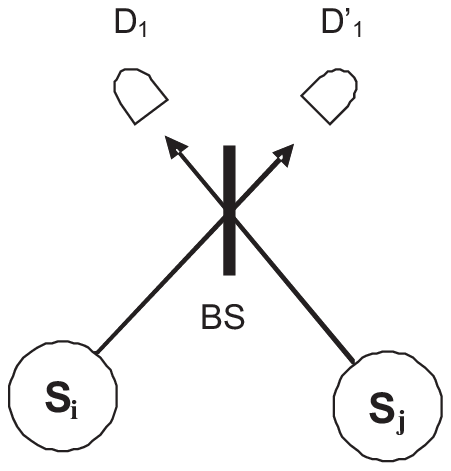}
\vspace*{3pt}
\fcaption{Elementary scheme with one beam-splitter BS.
The sources $S_{i}$, $S_{j}$ are chosen to be, in turn,
ot the same ($i=j=1,2$) and different ($i=1$, $j=2$) 
type.  Single click in either of the 
detectors $D$, $D'$ corresponds to projection into 
symmetric  two-qubit subspace of photon polarization, while
two clicks represent projection onto one-dimensional 
antisymmetric (singlet) subspace. 
\label{fig2}}
\end{figure}

The experimental setup is elementary. We have sources $S_{i}$, $S_{j}$,
where we put either $i=j=1,2$ (sources of the same type)
or, say $i=1$, $j=2$ (different sources) then we have a beamsplitter 
(in equal distance to the source) and two detectors behind it (see Fig. \ref{fig2}).
If two photons form sources $S_{i}$, $S_{j}$ meet  on the beamsplitter
and the two detectors click, we have so--called anticoalescence event, 
which happens  with probability $p^{(-)}_{ij}$  \cite{Bovino}.
Otherwise we deal with a coalescence result 
which occurs  with probability $p^{(+)}_{ij}= 1- p^{(-)}_{ij}$.
Putting all three probabilities of anti-coalescence into formula
(\ref{super-fidelity-3probabilities}) we reproduce the
expression for  super--fidelity.

This seems to be the most easy experiment with two sources to perform. 
Such an experiment can be realized for two sources of photons engineered with help of controlled 
decoherence (in a way similar to Ref. \cite{Kwiat})
corresponding to  two different mixed states of a qubit.

\begin{figure}[ht]
\centering
\includegraphics{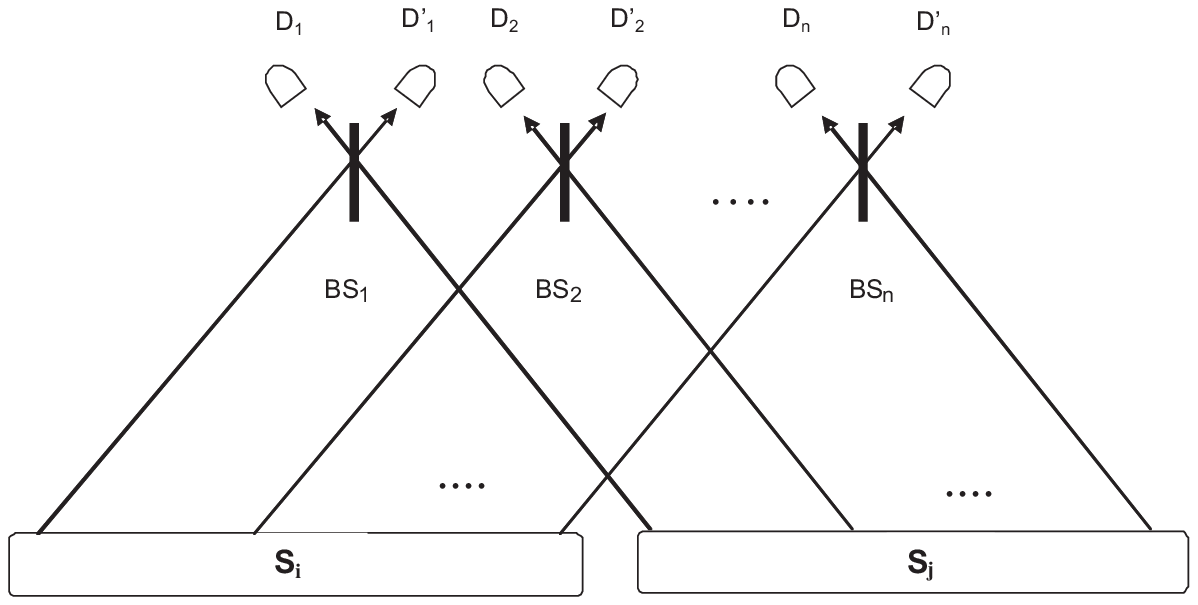}
\vspace*{3pt}
\fcaption{The scheme for measurement of super--fidelity of the states
of n-photon polarizations. According to (\ref{probabilities-n-photons}) only the
events with
double clicks in odd number of detector pairs contribute to each of the three probabilities 
in the formula (\ref{super-fidelity-3probabilities}). 
\label{fig3}}
\end{figure}

The above scheme immediately extends to the case of states $\rho_{1}$, $\rho_{2}$
are defined on $N=2^{n}$-dimensional Hilbert space representing polarization 
degrees of freedom of $n$ photons. 
In this case the total Hilbert space  is ${\cal H}_{N}=(\Cplx^{2})^{\otimes n}$
and the scheme reads
as in Fig. \ref{fig3} (compare \cite{Alves, ASH-Generalised-Entropies}).
If the probability  $p^{(s_k),k}_{ij}$ with $s_{k}=-1$ ($s_{k}=+1$) 
corresponds to anticoalescence (coalescence) on $k$-th beamsplitter,
i.e. it represents the probability of two clicks
(one click) in the pair of detectors $D_k$, $D_{k}'$, then the total probabilities:
\begin{equation}
p^{(-)}_{ij}=\sum_{s_{1},s_{2}...,s_{n}: \ s_{1}s_{2}...s_{n}=-1}p^{(s_1),1}_{ij}p^{(s_2),2}_{ij}...
p^{(s_n),n}_{ij}, \ \ \ i,j=1,2
\label{probabilities-n-photons}
\end{equation}
are these we put into  (\ref{super-fidelity-3probabilities}).
In the formula above we count all the cases when an odd number of anti-coalescence 
events occurs, 
provided that there is no photon losses during the experiment.

For $n=2$ this type of experiment has already been performed
with two two-photon sources producing entangled states \cite{Bovino}.
However, the sources were considered to provide the same state on average rather 
than two different ones. A similar reasoning was used in  another recent experiment,
in which photon polarization and momentum degrees of freedom 
were used to estimate the concurrence where additional strong assumption 
about purity of each copies were also used \cite{Walborn}.
In general, measurements schemes of quantities like purity, concurrence, sub-- and super--fidelity 
in the collective framework like the one presented here requires the assumption that
the sources producing states are stationary and memoryless.
Quite remarkably this is the same assumption one makes in quantum tomography. 
As discussed in \cite{vanEnk,vanEnkEtAl} there may be difficulties with satisfying it in real experimental scenarios for instance due to classical correlations between the consecutive copies of the system.
In other words the condition of having the global state in $\varrho^{N}$ form (which mathematically corresponds to quantum de Finnetti condition \cite{QuantumDeFinnetti}) may not be obeyed. This point requires more analysis, and leads, in general to nontrivial issues. It seems however that stationarity and memoryless character 
of the source can usually be satisfied approximately. Then the present measurement would serve
(similarily like quantum tomography does, though may be in a way more sensitive to source correlations) 
as an approximative, coarse-grained-like characteristic of the states sources 
under reasonable physical assumptions.
\subsection{Quantum networks}
There is yet another method of detection of quantities that may be considered here.
This is a method based on quantum networks. It is known that a
unitary operation $U$ acting on state $\sigma$ 
but {\it controlled} by  a qubit in the superposition state $|+\rangle\equiv 
\frac{1}{2}(|0\rangle +|1\rangle)$ reproduces the value $Re (\tr U \sigma)$ 
directly as a mean value of the Pauli matrix $\langle \sigma_{x}\rangle $
 measured on the controlled qubit \cite{Knill,EkertEtAl}.
This fact allows us to measure  certain nonlinear functions of the state.
To get  $\tr \rho^{k}$ one takes  $k$ copies of the state, 
$\sigma=\rho^{\otimes k}$ and takes for $U$ the operator of  
cyclic permutation, in full analogy to $V^{\pi}$ used in previous subsection.
To measure the overlap of two matrices, $\rho_1$, $\rho_2$, one 
takes $\sigma=\rho_1 \otimes \rho_2$ and uses the SWAP operator,  $U=V$.

 The corresponding network is already 
provided explicitly in Ref. \cite{EkertEtAl}, so we shall not write it down 
here. Such a  network allows one to measure all three quantities needed to 
reproduce the super--fidelity $G$. Indeed, the network produces {\it directly} 
(as mean values of $\sigma_z$ on controlled qubit)  
all three mean values: $\tr \rho_i\rho_j$, $i=1,2$
provided that states forming the input of the controlled part of the network 
are  $\rho_{i} \otimes \rho_{j}$. Some alternative constructions of programmable networks 
designed to measure super--fidelity are also possible.

\begin{figure}[ht]
\begin{center}
\[
\Qcircuit @C=0.53cm @R=0.45cm
{ &\dstick{\ket{\Psi_{12}}}  & &\qw  &\qw&\ctrl{2}&\qw&\ctrl{2}&\qw&\qw&\qw&\qw&\ctrl{2}&\qw&\qw&\qw&\qw&\qw&\\
  &                          & &\qw &\ctrl{2}&\qw&\qw&\qw&\qw&\ctrl{1}\qw&\qw&\ctrl{1}&\qw&\qw&\qw&\qw&\qw&\qw&\\
  &{\ket{0}} &  & \gate{X}&\qw&\targ&\ctrl{1}&\targ&\gate{X}&\targ&\ctrl{1}&\targ&\targ&\ctrl{1}&\ctrl{1}&\ctrl{1}&\qw&\qw&\\
  &{\ket{0}}        & & \gate{H}&\gate{\pi}&\qw&\ctrl{2}&\qw&\qw&\qw&\ctrl{4}&\qw&\qw&\ctrl{2}&\ctrl{3}&\ctrl{4}&\gate{\alpha}&\gate{H}&\measureD{M}&\\ 
  &{\rho_1}          &  & \qw &\qw&\qw&\qswap&\qw&\qw&\qw&\qw&\qw&\qw&\qswap&\qw&\qswap&\qw&\qw&\\
  &{\rho_2}         & & \qw &\qw&\qw&\qswap&\qw&\qw&\qw&\qw&\qw&\qw&\qswap&\qswap&\qw&\qw&\qw&\\
  &{\rho_1}         &  & \qw &\qw&\qw&\qw&\qw&\qw&\qw&\qswap&\qw&\qw&\qw&\qswap&\qw&\qw&\qw&\\
  &{\rho_2}         &  & \qw &\qw&\qw&\qw&\qw&\qw&\qw&\qswap&\qw&\qw&\qw&\qw&\qswap&\qw&\qw&
}
\]	
\end{center}
\vspace*{3pt}
\fcaption{Example of programmable network allowing to measure in particular 
the quantities $\tr(\rho_1\rho_2\rho_1\rho_2)$
and $\frac{1}{2}\tr(\rho_1\rho_2\rho_1\rho_2)-\tr(\rho_1\rho_2)^{2}$.
The state $|\Psi_{12}\rangle$ represents the program (see the main text).
Symbol $M$ corresponds to the measurement of  Pauli matrix $\sigma_{z}$. The
last phase gate usually is chosen to be $\alpha=0$ unless the interferometric
picture with visibility is needed (cf. Ref. \cite{EkertEtAl}). See \cite{NC00} 
for the description of quantum gates used in this circuit.}
	\label{fig4}
\end{figure}

Similarly,  sub--fidelity $E$ can also be estimated with a 
network-based experimental scheme. 
Following reasoning from Ref. \cite{ADH-Single-Witness}
one constructs the following programmable quantum network --
see Fig. \ref{fig4}. Depending on the program state $|\Psi_{12}\rangle$ 
as a mean value of $\sigma_{z}$ of the measured controlling qubit
one gets
\begin{enumerate}
 \item[(i)] $\tr \rho_1\rho_2$ if $|\Psi_{12}\rangle=|0\rangle|0\rangle$,
 \item[(ii)] $\tr \rho_1\rho_2\rho_1\rho_2$ if
$|\Psi_{12}\rangle=|1\rangle|0\rangle$,
 \item[(iii)] $\frac{1}{2}\left(\tr \rho_1\rho_2\rho_1\rho_2 - (\tr
\rho_1\rho_2)^{2}\right)$ if $|\Psi_{12}\rangle =
(|0\rangle|1\rangle+|1\rangle|0\rangle)/\sqrt{2}$ \ {\sl i.e.} if it is in Bell state.
\end{enumerate}

The last quantity up to the factor $(-\frac{1}{4})$ is just the 
quantity that occurs under the square root in the formula for $E$.
In general to estimate  the sub--fidelity $E$ we 
may ''run'' the first ''program'' and either the second or the third one.
Alternatively, we may run all three programs and use the data
to verify the accuracy of the experiment by comparing  the two 
partially independent estimates of $E$ obtained in that way.
It is easy tu see, that the same network can be used to estimate super--fidelity $G$ 
if  one  puts as an input $\rho_{i} \otimes \rho_{i} \otimes \rho_{j}\otimes \rho_{j}$, 
$j=1,2$. 
\section{Distances  and geometry of the space of states}

\subsection{Hilbert-Schmidt distance and flat geometry}

The geometry of the space of quantum states depends on the metric used
\cite{BC94,PS96,ZS01,BZ06}.
The set ${\states}_N$ of mixed states of size $N$
reveals the Euclidean (flat) geometry if
it is analyzed with respect to the {\sl Hilbert-Schmidt distance},
\begin{equation}
D_{HS}(\rho_1,\rho_2)=\sqrt{ \tr   [(\rho_1 - \rho_2)^2] }.
  \label{HilbSchmi}
\end{equation}

To demonstrate this property let us first concentrate on the simplest case,
$N=2$.
Making use of the notion of a coherence vector $\vec \tau$
any state of a qubit can be written in the {\sl Bloch representation}
\begin{equation}
\rho =  \frac {\1}{N} +
{\vec \tau}\cdot {\vec \lambda} .
\label{Pauli}
\end{equation}
Here  $\vec{\lambda}$ denotes the vector of
three rescaled traceless Pauli matrices
$\{\sigma^x, \sigma^y, \sigma^z \}/{\sqrt{2}}$,
which are orthogonal in the sense of the
Hilbert-Schmidt scalar product,
$\langle \lambda^k |\lambda^m\rangle=
\tr(\lambda^k)^{\dagger}\lambda^m=\delta_{km}$.
Together with $\lambda^0={\1}/\sqrt{2}$ they
form an orthonormal basis in the space of complex density matrices
of size two. Due to Hermiticity of $\rho$
the three-dimensional Bloch vector $\vec \tau$ is real.
Positivity condition implies
$|\vec \tau|\le 1/\sqrt{2}=R_2$ with equality for pure states,
which form the Bloch sphere of radius $R_2$.
Representation (\ref{Pauli}) implies that
for any state of a qubit $\tr \rho^2=1/2 + |\tau|^2$.

Consider two arbitrary density matrices
and express their difference $\rho_1-\rho_2$
in the Bloch form.
The entries of this difference consist of the differences
between components of both Bloch vectors ${\vec \tau}_1$  and
${\vec \tau}_2$. Therefore Hilbert-Schmidt distance
induces the flat (Euclidean) geometry of ${\states}_2$,
\begin{equation}
  D_{HS}\bigl( \rho_{{\vec \tau}_1}, \rho_{{\vec \tau}_2}\bigr)=
  D_{E}({\vec \tau}_1, {\vec \tau}_2) ,
\label{densHS}
\end{equation}
where $D_E$ is the Euclidean distance between both
Bloch vectors in ${\mathbbm R}^3$.

It is worth to add that expression (\ref{densHS})
holds for an arbitrary $N$. In this case
$\vec \tau $ is a real vector with $N^2-1$ components,
while the vector ${\vec \lambda}=\{\lambda^k\}_{k=1}^{N^2-1}$  in (\ref{Pauli})
denotes the set of $N^2-1$ traceless generators of
the group $SU(N)$. Positivity of $\rho$ implies
that the length of the Bloch vector
is limited by
\begin{equation}
  |{\vec \tau}| \le D_{HS}({\1}/N,|\psi\rangle \langle \psi|) =
\sqrt{ \frac{N-1}{N}} =: R_N .
\label{radiusN}
\end{equation}
For $N=2$ the condition $|{\vec \tau}| \le R_2$
is sufficient to imply that the corresponding matrix is positive and
represents a state, while for $N\ge 3$ it is only a necessary condition \cite{BZ06}.
This is related to the fact that with respect to  the flat, H--S geometry
 the set  ${\states}_2$ forms a full $3$-ball,
while for larger $N$ the set
${\states}_N$ forms a convex subset of the $(N^2-1)$-dimensional ball of
radius $R_N$ centered at $\rho_*={\1}/N$.

\subsection{Bures distance and the geometry it induces}

The notion of fidelity, introduced in (\ref{fidel}), can be used to
define  the {\sl Bures distance} \cite{Bu69,Uh76}
 \begin{equation}
  D_{F}(\rho_1,\rho_2)   =   \sqrt{ 2 - 2 \sqrt{F(\rho_1,\rho_2)}} .
 \label{Bures}
\end{equation}
or the {\sl Bures length} \cite{Uh95}
(later called {\sl angle} in \cite{NC00}),
\begin{equation}
D'_F (\rho_1,\rho_2)  :=  {\rm arccos}
\sqrt{F (\rho_1,\rho_2)}=\frac{1}{2}{\rm arccos}
\Bigl( 2F(\rho_1,\rho_2)-1\Bigr) .
\label{anglefid}
\end{equation}
For any pair of pure states the Bures length  coincides
 with their Fubini--Study distance,
$D'_F\bigl(\rho_{\psi}, \rho_{\phi}\bigr)=
 d_{FS}\bigl( |\psi\rangle,|\phi\rangle\bigr)=
 {\rm arccos}|\langle \psi|\phi\rangle|$.

The Bures metric is distinguished by its rather special properties:
it is a Riemannian, monotone metric \cite{Pe96},
Fisher adjusted metric \cite{PS96},
closely related to the statistical distance \cite{BC94}.

It is not difficult to describe the geometry of the set
of mixed states of a single qubit induced by the Bures metric.
Consider a mixed state $\rho\in {\states}_2$
 and its  transformation  proposed in  \cite{Uh92}
\begin{equation}
\rho(x,y,z) \to \Bigl( x,y,z,t=\sqrt{\frac{1}{2}-x^2-y^2-z^2} \Bigr) .
\label{blow1}
\end{equation}
It blows up the Bloch ball ${\bf B}^3$ of radius $R_2=1/\sqrt{2}$
into a hyper-hemisphere $\frac{1}{2}S^3$ of the same radius.
The original variables $(x,y,z)$ denote the
parameters of the state in the Bloch vector representation.
The auxiliary variable reads
$t=\sqrt{1/2-|\tau|^2}$ in terms of the Bloch vector,
so that  $t^2+\tr \rho^2=1$.
The maximally mixed state $\rho_*=(0,0,0)$, is mapped into a hyper-pole.
It is equally distant from all pure states located at the hyper-equator $S^2$,
which form the boundary of ${\states}_2$.

Any state $\rho$ is uniquely represented by an
'extended Bloch vector',
${\vec v}=(x,y,z,t)$ of length $R_2$.
The auxiliary variable reads
$t=\sqrt{1/2-|\tau|^2}$ in terms of the Bloch vector,
so that  $t^2+\tr \rho^2=1$.
Consider two  states $\rho_1$ and $\rho_2$,
described by
two vectors $\vec v_1$ and ${\vec v_2} \in {\mathbbm R}^4$,
which form the angle  $\vartheta$.
Since for any one-qubit states the bound (\ref{eqn:main-theorem}) becomes an
equality,
we see that fidelity between them reads
$F(\rho_1,\rho_2)=\tr \rho_1\rho_2+\sqrt{t_1^2 t_2^2}
=1/2+{\vec \tau_1} \cdot {\vec \tau_2}+t_1 t_2$.
This can be rewritten with the use of extended vectors $\vec v_i$
and the angle between them,
$F=1/2+{\vec v_1}\cdot{\vec  v_2}=1/2+R_2^2 \cos \vartheta$.
Since $R_2^2=1/2$  we find
\begin{equation}
\vartheta =\arccos\bigl( 2F-1\bigr) =
2D'_F (\rho_1,\rho_2) ,
\label{theta}
\end{equation}
which shows that the Bures length (\ref{anglefid})
between any two mixed states is
proportional to the Riemannian distance between
the corresponding points at the Uhlmann hemisphere.

Making use of the fidelity $F$ one can also define 
other distances in the space of quantum states.
 For instance, Gilchrist et al. \cite{GLN05} have shown
 that the {\sl root infidelity}
\begin{equation}
  C(\rho_1,\rho_2)   =   \sqrt{ 1 - F(\rho_1,\rho_2)} 
 \label{infidelity}
\end{equation}
satisfies the triangle inequality and thus
introduces a metric. This very quantity can be used to 
bound the trace distance 
$D_{\rm tr}(\rho_1,\rho_2)=
\frac{1}{2}{\rm Tr} |\rho_1-\rho_2| \le C(\rho_1,\rho_2)$
from above \cite{FG99,BZ06}.
Note that the Bures distance, Bures length
and root infidelity are functions of the same quantity,
so they generate the same topology.

\subsection{Modified Bures length}

In analogy to (\ref{Bures})  and (\ref{anglefid}) one may ask
whether
 \begin{equation}
  D_{G}(\rho_1,\rho_2)  =   \sqrt{ 2 - 2 \sqrt{G(\rho_1,\rho_2)}} .
 \label{Buresup}
\end{equation}
and
\begin{equation}
D'_G (\rho_1,\rho_2)  :=  {\rm arccos} \sqrt{G (\rho_1,\rho_2)}
\label{angleup}
\end{equation}
define distances. This is obvious for $N=2$ because $F=G$.
The situation changes for $N\ge 3$,
for which the $D_F$ and $D_G$ do differ and only $F \le G$
is valid.

We do not know, whether $D_G$ and $D'_G$ are distances.
However, it can be proved that a direct analogue
of the  root infidelity (\ref{infidelity})
\begin{equation} 
\label{new.1a}
C'(\rho_1, \rho_2) = \sqrt{ 1 -  G(\rho_1,\rho_2)}
\end{equation}
is a genuine distance. 
 The same is true for the {\sl modified Bures length}, 
\begin{equation} \label{new.1b}
D'_M( \rho_1, \rho_2) = \arccos G(\rho_1, \rho_2)  .
\end{equation}
\proof{  Let us call ${\cal L}$ the direct sum of the real linear
space of all Hermitian operators and the 1-dimensional space
of real numbers. Its elements are $\{H, x\}$, $H$ Hermitian,
$x$ a real number. ${\cal L}$ becomes Euclidean (i.e. a real
Hilbert space) by defining the scalar product
\begin{equation} \label{new.2}
(\{H_1, x_1\} , \{H_2, x_2\}) = \tr H_1 H_2 + x_1 x_2 .
\end{equation}
Let us denote by $B({\cal L})$ the unit ball of ${\cal L}$ and
by $S({\cal L})$ the unit sphere. Our proof rests on the embedding
of the Hermitian operators
\begin{equation} \label{new.3a}
B_N = \left\{ H  | \ \tr H = 1, \tr H^2 \leq 1 \right\}
\end{equation}
into $S({\cal L})$ by
\begin{equation} \label{new.3b}
H  \to  \xi_H := \left\{ H, \sqrt{1 - \tr H^2} \right\}  .
\end{equation}
Clearly, $(\xi_H, \xi_H) = 1$, and from (\ref{new.2}) we get
\begin{equation} \label{new.3c}
(\xi_H, \xi_{H'}) = G(H, H')  .
\end{equation}
Now it is obvious that $\sqrt{2-2G}$ is the Euclidean distance
between $\xi_H$ and $\xi_H'$, provided $H$ and $H'$
belong to $B_N$. Because the density operators form a subset
of $B_N$, (\ref{new.1a}) is a distance.

From (\ref{new.3c}) we get $G(H, H') = \cos \alpha$,
where $\alpha$ is the angle from which $\xi_H$ and $\xi_H'$
are seen from the center of the ball $B({\cal L})$. Thus,
$\arccos G = \alpha$ and, in particular, (\ref{new.1b}) is a distance.
}

Let us now return to the two conditions of
(\ref{new.3a}). They are equivalent with
\begin{equation} \label{new.4}
B_N = \left\{  H  | \ \tr H = 1, 
\tr (H - \frac{1}{N} \1)^2 \leq
\frac{N-1}{N}  \right\}
\end{equation}
and they describe the smallest ball containing the state
space. $B_N$ is an affine translate by $1/N$ of the
generalized Bloch-ball, \cite{NC00}. $B_N$ is centered at
$A = N^{-1} \1$ and is of radius $\sqrt{(N-1)/ N}$.

Above we have embedded $B_N$ by the map (\ref{new.3b})
into the sphere $S({\cal L})$.  Just this gives the opportunity
to apply Mielnik's definition \cite{Mi74} for a
transition probability (he also called it {\sl affine ratio})
of two extremal states of a compact convex set.
In our case the compact convex set is
$B({\cal L})$ and its extremal part is $S({\cal L})$.
At the case at hand, Mielnik's procedure starts with
first choosing an extremal point $\xi \in S({\cal L})$
and selecting all affine functions $l$ satisfying
$l(\xi) = 1$ and  $0 \leq l \leq 1$  on $B({\cal L})$.
Any such function can be written
\begin{equation} \label{new.5a}
l(\eta) =
\frac{a + (\xi, \eta) + 1}{a + 2 } ,
\end{equation}
with $a \geq 0$ and $\eta \in {\cal L}$ arbitrarily.
Now we have to vary over all these affine functions,
\begin{equation} \label{new.5b}
p_{M}(\eta, \xi) := \min_l l(\eta)
= \min_a \frac{a + (\xi, \eta)}{2 + a} ,
\end{equation}
to get Mielnik's transition probability
\begin{equation} \label{new.5c}
p_{M}(\xi, \eta) = \frac{1 + (\xi, \eta)}{2} .
\end{equation}
Returning to $H, H' \in B_N$, we can write
\begin{equation} \label{new.5d}
p_M(H, H') := p_M(\xi_H, \xi_{H'}) = \frac{1 + G(H, H')}{2}
\end{equation}
and, in becoming even more special by choosing two density
operators for $H$ and $H'$ in the equation above, we arrive at
\begin{equation} \label{new.5e}
D_{M}(\rho_1, \rho_2) = 2 \sqrt{1 - p_M(\rho_1, \rho_2)}
= 2 \sin \frac{\alpha}{2},
\end{equation}
(using $2 \cos^2 = 1 + \cos$) and also at
\begin{equation} \label{new.5g}
D'_{M}(\rho_1, \rho_2) = 2 \arccos \sqrt{p_M(\rho_1, \rho_2)}  .
\end{equation}

\section{Concluding remarks}

In this paper we analyzed various bounds for quantum fidelity. Two quantities,
we propose to call sub-- and super--fidelity, posses particularly nice
properties. On one hand these quantities form
universal lower and upper bounds for
the fidelity. Moreover,  with respect to the tensor product
they display sub-- and super--multiplicativity.

On the other hand, quantities $E$ and $G$
are much easier to calculate than the original fidelity $F$.
To compute any of these bounds it is enough to evaluate three traces only.
Thus one can expect,  the quantities introduced in this paper
might become useful for various tasks of
the theory of quantum information processing.
Furthermore, under a realistic assumption
that several copies of both states are available,
it is possible to design a scheme to measure experimentally
sub-- and super--fidelity between arbitrary mixed states.
For instance, the measurement of  super--fidelity 
is possible if one has three copies of each state.
In this paper  we have worked out concrete schemes
of such experiments concerning the super--fidelity between
any two mixed states
representing the polarization of photons.

\nonumsection{Acknowledgements}
\noindent
We would like to thank A. Buchleitner for inviting three of us
to Dresden in September 2005 for a workshop on Quantum Information,
during which our collaboration on this project was initiated.
It is also a pleasure to thank I. Bengtsson and  M. Horodecki 
for inspiring discussions. 
J.A.M. would like to thank Iza Miszczak for her help.

We acknowledge financial support by the Polish
Ministry of Science and Higher Education under the grants number N519 012
31/1957 and  DFG-SFB/38/2007, by the LFPPI network and by 
the European Research Project SCALA.

\nonumsection{Note added}
\noindent After this paper was submitted we learned about
a related work by Mendonca et al. \cite{MNMFL08}
in which the super--fidelity was independently introduced
and was called an 'alternative fidelity' measure.
In this valuable work the authors provide an alternative proof of
super--multiplicativity of $G$, discuss its relation to the trace distance
and analyze the distance $G$ induces into the space of mixed
quantum states,
and prove that $G$ is {\sl jointly} concave in its two arguments.

\nonumsection{References}


\appendix{~~~Algebraic  facts}\label{sec:app}

\noindent In this appendix we collect useful algebraic facts,
which are used in the main body of the paper.

\begin{fact}[From corollary IX.5.3 in \cite{bhatia}]\label{fact:2}
For any positive matrices $A$ and $B$ and every unitarily invariant norm
$|||\cdot|||$ we have
\begin{equation}
|||A^\nu B^{1-\nu}||| \leq|||A|||^\nu |||B|||^{1-\nu},
\end{equation}
where $\nu \in [0,1]$.
\end{fact}

\begin{fact}[From corollary IX.5.4 in \cite{bhatia}]\label{fact:3}
For any positive matrices $A$ and $B$ and every unitarily invariant norm
$|||\cdot|||$ we have
\begin{equation}
|||A^\nu B^\nu||| \leq ||| \1 |||^{1 -\nu} |||AB|||^{\nu},
\end{equation}
where $\nu \in [0,1]$.
\end{fact}

Next two facts can be found in \cite{coope}.
\begin{fact}\label{fact:4}
  Matrix $AB$ is similar to matrices $\sqrt{A}B\sqrt{A}$ and
  $\sqrt{B}A\sqrt{B}$.
\end{fact}

\begin{fact}\label{fact:5}
  For positive matrices $A$ and $B$ matrix $AB$ has positive eigenvalues.
\end{fact}

\begin{fact}\label{fact:6}
If $p_1 + p_2 + \dots + p_n = 1$ and $p_i \geq 0$ then
\begin{equation}
  1 - p_1^2 - p_2^2 - \dots -p_n^2 = \sum_{i \neq j} p_i p_j.
\end{equation}
\end{fact}
%

\begin{prop}\label{prop:ProdStrongIso}
Let $g$ be defined as
\begin{equation}
  g(x) = \sum_{i\not=j}\sqrt{x_i}\sqrt{x_j} .
\end{equation}
For $x , y \in \Real_+^n$ such that
\begin{equation}\label{def:weakProdMaj}
\prod_{i=1}^k x_i \leq \prod_{i=1}^k y_i , \ \mathrm{ for } \  k=1,\dots,n ,
\end{equation}
with equality for  $k=n$, we have
\begin{equation}
g(x) \leq g(y).
\end{equation}
\end{prop}
\proof{
We introduce notation
\begin{equation}
g_i(\cdot) = \frac{\partial g}{\partial x_i} (\cdot) .
\end{equation}
Direct computation shows that function $g$ satisfies
\begin{equation}\label{eqn:ProdSchurOstrowski}
u_1 g_1(u) \geq u_2 g_2(u) \geq \dots \geq u_n g_n(u) ,
\end{equation}
for $u \in \Real^n$ such that $u_1 \geq u_2 \geq \dots \geq u_n \geq 0$.
We denote $\alpha_i = \log(x_i)$ and $\beta_i = \log(y_i)$.
Note that (\ref{def:weakProdMaj}) can be rewritten as
\begin{equation}
\sum_{i=1}^k \alpha_i \leq \sum_{i=1}^k \beta_i  \text{ for } k=1,\dots,n, 
\end{equation}
with equality for $k=n$.
We define new function
\begin{equation}\label{def:funkcjaH}
h(v) = g(e^{v_1},e^{v_2},\dots, e^{v_n}).
\end{equation}
For a given vector $u$ such that $u_1 \geq u_2 \geq \dots \geq u_n \geq 0$, and
$v_i = \log(u_i)$ we have
\begin{equation}
v_1 \geq v_2 \geq \dots \geq v_n .
\end{equation}
Using (\ref{eqn:ProdSchurOstrowski}) we can  write
\begin{equation}
e^{v_1} g_1(e^{v_1},e^{v_2},\dots, e^{v_n}) \geq \dots \geq e^{v_1}
g_n(e^{v_1},e^{v_2},\dots, e^{v_n}) .
\end{equation}
Now from above and (\ref{def:funkcjaH}) we have
\begin{equation}
h_1(v) \geq h_2(v)  \geq \dots \geq h_n(v).
\end{equation}
Note now that function $h$ satisfies condition from
\cite[Theorem 3.6]{buliga}
and thus it is {\it Schur-convex}, so
\begin{equation}
h(\alpha) \leq h(\beta).
\end{equation}
Using (\ref{def:funkcjaH}) we can write
\begin{equation}
g(x) \leq g(y).
\end{equation}
Thus the proof is complete.
}

\begin{fact}[H\"older's inequality \cite{handbook}]\label{fact:holder}
  For $ a >1, b = a/(a-1)$  and positive semidefinite $A$ and $B$ we have
  \begin{equation}
    \tr (A B)  \leq  (\tr A^a)^{1/a} (\tr B^b)^{1/b}.
  \end{equation}
\end{fact}

\begin{fact}\label{fact:inequality1}
  For density matrices $A$ and $B$ we have
  \begin{equation}
    1-\sqrt{\tr A^2}\sqrt{\tr B^2} \geq \sqrt{1-\tr A^2}\sqrt{1-\tr B^2} .
  \end{equation}
\end{fact}
\proof{
This inequality can be rewritten in equivalent form
\begin{equation}
  1-2\sqrt{\tr A^2 \tr B^2} + \tr A^2 \tr B^2 \geq 1 - \tr A^2 - \tr B^2 +
  \tr A^2 \tr B^2 ,
\end{equation}
which is equivalent to
\begin{equation}
  \sqrt{\tr A^2 \tr B^2}  \leq  \frac{\tr A^2 + \tr B^2}{2}.
\end{equation}
This  completes the proof since for any positive numbers the arithmetic mean is
always greater than or equal to the geometric mean.
}
\begin{fact}[Maclaurin inequality \protect{\cite[p. 5]{Biler}}]\label{fact:Maclaurin}
For a given matrix $A$ of rank $r$ and with $r$ positive eigenvalues we have
  \begin{equation}
    \sqrt[k]{\frac{s_k(A)}{\binom{r}{k}}} \geq \sqrt[k+1]{\frac{s_{k+1}(A)}{\binom{r}{k+1}}}
  \end{equation}
for $1 \leq k < r$.
\end{fact}
\appendix{~~~Proof of the lower bound (\ref{lowerbis})}

\noindent To prove that sub--fidelity $E$ is not larger than fidelity $F$,
let us take a look at equations
(\ref{Fs1}) and (\ref{sub-s2})
in which both quantities are expressed in terms of the second symmetric
function. We can rewrite the function $s_2$, which forms fidelity,
\begin{eqnarray}
s_2(\sqrt{A^{1/2} B A^{1/2}})
&=& \sum_{i<j} \lambda_i(\sqrt{A^{1/2} B A^{1/2}}) \lambda_j(\sqrt{A^{1/2} B
A^{1/2}}) \\
&=& \sum_{i<j} \sqrt{\lambda_i(A^{1/2} B A^{1/2})} \sqrt{\lambda_j(A^{1/2} B
A^{1/2})} \\
&=& \sum_{i<j} \sqrt{\lambda_i(A B )} \sqrt{\lambda_j(A B)} .
\end{eqnarray}
The last equality is the consequence of similarity of matrices
$A^{1/2} B A^{1/2}$  and $A B$.
Making use of subadditivity of square root we obtain
\begin{eqnarray}
s_2(\sqrt{A^{1/2} B A^{1/2}})
&\geq& \sqrt{\sum_{i<j} \lambda_i(A B ) \lambda_j(A B) } \\
&=& \sqrt{s_2(A B)}.
\end{eqnarray}
As a consequence we get
\begin{equation}
F(A,B) = \tr AB + 2 s_2(\sqrt{A^{1/2} B A^{1/2}}) \geq
\tr AB + 2 \sqrt{s_2(A B)} = E(A,B).
\end{equation}

\appendix{~~~Proof of the lower  bound  (\ref{determin}) }
  \label{sec:app2}
\noindent To prove inequality (\ref{determin}) we use Fact \ref{fact:Maclaurin} (Maclaurin inequality) 
and obtain 
\begin{equation}
\left( \frac{s_2\left(\sqrt{A^{1/2} B A^{1/2}} \right)}{\binom{r}{2}} \right)^{1/2} \geq 
\left( \frac{s_{r}\left(\sqrt{A^{1/2} B A^{1/2}} \right)}{\binom{r}{r}}\right)^{1/r}.
\end{equation}
Using Fact \ref{fact:4} we get
\begin{eqnarray*}
s_2\left(\sqrt{A^{1/2} B A^{1/2}} \right) &\geq&
\binom{r}{2} \left( {s_{r}\left(\sqrt{A^{1/2} B A^{1/2}} \right)}\right)^{2/r}  
=
\binom{r}{2} \left({\prod_{i=1}^r \lambda_{i}\left(\sqrt{A^{1/2} B A^{1/2}} \right)} \right)^{2/r} \\
&=&
\binom{r}{2} \left({\prod_{i=1}^r \sqrt{\lambda_{i}\left(A^{1/2} B A^{1/2} \right)}}\right)^{2/r}  
=
\binom{r}{2} \left({\prod_{i=1}^r \lambda_{i}\left(A B \right)} \right)^{1/r}\\
&=&
\binom{r}{2} \sqrt[r]{s_r(AB)}.
\end{eqnarray*}
Now using (\ref{Fs1}) we write 
\begin{equation}
F(A,B) = \tr AB  + 2 s_2\left(\sqrt{A^{1/2} B A^{1/2}} \right) \geq
\tr AB  + r(r-1)  \sqrt[r]{s_r(AB)}.
\end{equation}
\halmos


\appendix{~~~Proofs of Lemmas}\label{sec:LemmaProof}

\noindent \proofof{Lemma \ref{lemma:s2(X)<s2(d(A)d(B))}}{
Observe that the matrix $\rho_1^{1/2} \rho_2 \rho_1^{1/2}$ is similar to
$\rho_1 \rho_2$ and thus
\begin{eqnarray*}
2 s_2 \left(\sqrt{\rho_1^{1/2}\rho_2 \rho_1^{1/2}} \right)
&=& \sum_{i \neq j} \lambda_i \left(\sqrt{\rho_1^{1/2}\rho_2 \rho_1^{1/2}}
\right) \lambda_j\left(\sqrt{\rho_1^{1/2}\rho_2 \rho_1^{1/2}} \right)
\\
&=& \sum_{i \neq j} \sqrt{ \lambda_i \left(\rho_1^{1/2}\rho_2 \rho_1^{1/2}
\right) \lambda_j\left(\rho_1^{1/2}\rho_2 \rho_1^{1/2} \right) } 
\\
&=& \sum_{i \neq j} \sqrt{ \lambda_i \left(\rho_1 \rho_2  \right)
\lambda_j\left(\rho_1 \rho_2  \right) },
\end{eqnarray*}
where $\lambda_i(A)$ denotes $i^{\text{th}}$ eigenvalue of a matrix $A$.

Let us define a function $g : R^n \to R$ which acts on a vector $\vec x$
of non-negative numbers
\begin{equation}
  g({\vec x}) := \sum_{i\not=j}\sqrt{x_i x_j} .
\end{equation}
It allows one  to rewrite
\begin{equation}\label{eqn:s_2-as-g}
  2 s_2\left( \sqrt{\rho_1^{1/2}\rho_2 \rho_1^{1/2}} \right) = g \bigl( {\vec
\lambda}(\rho_1\rho_2) \bigr) ,
\end{equation}
where $\vec \lambda(A)$ denotes the vector
of eigenvalues of $A$.

From \cite[Theorem 3.3.2 and 3.3.4]{hj2} we obtain
\begin{equation}
\prod_{i=1}^k \lambda_i(\rho_1\rho_2) \leq \prod_{i=1}^k
\lambda_i(\rho_1)\lambda_i(\rho_2) \text{ for } k=1,\dots,n ,
\end{equation}
with equality for $k=n$.
Making use of
Proposition \ref{prop:ProdStrongIso} from Appendix A
with
$x_i = \lambda_i(\rho_1 \rho_2)$ and $y_i = \lambda_i(\rho_1)
\lambda_i(\rho_2)$
we obtain
\begin{equation}
g\bigl( {\vec \lambda} (\rho_1\rho_2) \bigr)  \leq
g\bigl( {\vec \lambda (\rho_1)} \circ  {\vec \lambda(\rho_2)} \bigr),
\end{equation}
where $\circ$ denotes Hadamard product, \cite[Definition 7.5.1]{hj2}.
Now making use of (\ref{eqn:s_2-as-g}) we get
\begin{equation}
  s_2\left( \sqrt{\rho_1^{1/2}\rho_2 \rho_1^{1/2}} \right) 
  \leq 
  s_2\left(\sqrt{ \diag({\vec \lambda(\rho_1)}) \diag({\vec \lambda(\rho_2)})}\right).
\end{equation}
And thus the proof is complete.
}


\proofof{Lemma \ref{lemma:s2<sqrt(s2)}}{
For given density matrices  $\rho_1, \rho_2$ with eigenvalues $p_1, \dots ,
p_n$ and  $q_1, \dots , q_n$ respectively.
We denote diagonal matrices with entries on diagonal $p_1, \dots , p_n$ and
$q_1, \dots , q_n$ as $\diag(p),\diag(q)$ respectively.

Rewriting the second elementary function $s_2$ we obtain
\begin{equation}
2 s_2 \left(\sqrt{\diag(p) \diag(q)} \right) = \sum_{i \neq j}
 \sqrt{p_i q_i} \sqrt{p_j q_j}.
\end{equation}
On the other hand
\begin{equation}
2 \sqrt{s_2(\rho_1) s_2(\rho_2)} =
\sqrt{\left(1 - \sum p_i^2 \right) \left(1 - \sum q_i^2 \right)} .
\end{equation}


Let us define vectors $x, y \in \Real^{n^2}$
\begin{equation}
  x_{i,j} = \sqrt{p_i p_j} (1- \delta_{i,j}),\ y_{i,j} = \sqrt{q_i q_j} (1-
  \delta_{i,j}),
\end{equation}
where $x_{i,j} = x_{n(i-1) + j}$. 
Using Cauchy--Schwarz inequality 
 \begin{equation}
   |\scalar{x}{y}| \leq \sqrt{\scalar{x}{x}}\sqrt{\scalar{y}{y}},
 \end{equation}
we get 
\begin{equation}
\label{eq8}
\sum_{i \neq j} \sqrt{p_i q_i} \sqrt{p_j q_j} \leq \sqrt{\left(1 - \sum p_i^2
\right) \left(1 - \sum q_i^2 \right)}.
\end{equation}
This completes the proof.
}


\appendix{~~~The case $N=3$}\label{sec:app3}

\noindent In this section we are going to study the fidelity of two states
$\rho_1$ and $\rho_2$ in the case
where the rank $r$ of their product $\rho_1\rho_2$
is not greater than $3$.
As in Section \ref{sec:fid1} we will
denote eigenvalues of  $\sqrt{\rho_1^{1/2} \rho_2 \rho_1^{1/2}}$ by
$\lambda_i$,
so eigenvalues of $\rho_1^{1/2} \rho_2 \rho_1^{1/2}$ are given by
$\{\lambda^2_i\}$
and by similarity we have that the eigenvalues of $\rho_1 \rho_2$ are also given by
$\{\lambda^2_i\}$.
Since $r\le 3$, not more than three eigenvalues
of $\rho_1\rho_2$ are positive, so
the  third symmetric function (\ref{s33})
reads $s_3(\rho_1\rho_2)=(\lambda_1 \lambda_2 \lambda_3)^2$.
This is so for any two states of a qutrit,
so for $N=3$ one has $s_3(\rho_1\rho_2)={\rm det}(\rho_1\rho_2)$.

Consider now the expression for fidelity (\ref{eqn:fidelity-sum-singular})
which can be rewritten with the use of the second  symmetric function,
\begin{equation}
\label{fid_s2}
  F(\rho_1,\rho_2)    =  \tr \rho_1\rho_2  +
2  s_2 \Bigl( \sqrt{\rho_1^{1/2}\rho_2 \rho_1^{1/2}} \Bigr).
\end{equation}

The square of the symmetric function presented in
the above equation,
can be written as
$\bigl(\sum_{i<j} \lambda_i \lambda_j \bigr)^2 =
\sum_{i<j} \lambda_i ^2 \lambda_j^2 +R$.
The reminder $R$, defined implicitly by this equation,
is equal to zero if $r \le 2$
and the sum consists of a single term only.
It is difficult to handle $R$ generally. But if $r=3$
one has
\begin{equation}
\label{fid_R}
R=\lambda_1\lambda_2(\lambda_2\lambda_3+\lambda_3\lambda_1) + ... \ =
2(\lambda_1\lambda_2 \lambda_3)(\lambda_1+\lambda_2+\lambda_3) .
\end{equation}
However, in the particular case  $r\le 3$
discussed here used to (\ref{rootfid}) one has
 $\lambda_1+\lambda_2+\lambda_3=\sqrt{F}$
 while $\lambda_1 \lambda_2 \lambda_3=\sqrt{s_3(\rho_1\rho_2)}$.
Combining this with (\ref{eqn:fidelity-sum-singular})
we get the equation for fidelity satisfied for $r \le 3$
\begin{equation}
F = \tr \rho_1 \rho_2 + 2 \sqrt{ s_2(\rho_1\rho_2) +  2 \sqrt{F} \sqrt{s_3( \rho_1 \rho_2) }} .
\label{rank3}
\end{equation}
In the case $r \le 2$ the third function $s_3$ vanishes, so
this equation leads to  an expression,
$F  = \tr \rho_1 \rho_2 +    2 \sqrt{ s_2(\rho_1\rho_2)}=
 \tr \rho_1 \rho_2 +    2 \lambda_1\lambda_2$,
already discussed in Section \ref{sec:fid1}.

Another relation for fidelity is due to the fact that an assumption $r\le 3$
implies that
\begin{equation}
\sum_{j< k} \lambda_j \lambda_k =
(\lambda_1\lambda_2 \lambda_3)(\lambda_1^{-1}+\lambda_2^{-1}+\lambda_3^{-1}) .
\end{equation}
Therefore in this case one has
\begin{equation} \label{r3_s2}
s_2 \Bigl( \sqrt{A_1^{1/2} B A^{1/2} } \Bigr) =
\sqrt{ {\rm det}(AB) F(1/A,1/B)} .
\end{equation}
Lifting for a moment the assumption that the arguments of fidelity have to be
normalized,  we arrive therefore at another equation for fidelity
satisfied for $r\le 3$,
 \begin{equation}
    F (A,B) = \tr AB  +
            2\sqrt{ {\rm det}(AB) F(1/A,1/B) } .
    \label{rank3b}
  \end{equation}

\end{document}